\documentclass[12pt]{article}

\usepackage{amsmath}
\usepackage{amssymb}
\usepackage{amsthm}
\usepackage{latexsym}
\usepackage{color}
\usepackage{graphicx}
\usepackage{color}

\DeclareSymbolFont{calletters}{OMS}{cmsy}{m}{n}
\DeclareSymbolFontAlphabet{\mathcal}{calletters}

\def\be{\begin{eqnarray}}
\def\ee{\end{eqnarray}}

\def\b*{\begin{eqnarray*}}
\def\e*{\end{eqnarray*}}

\def\reff{\eqref}

\newtheorem{Theorem}{Theorem}[section]

\newtheorem{Proposition}[Theorem]{Proposition}

\newtheorem{Assumption}[Theorem]{Assumption}

\newtheorem{Lemma}[Theorem]{Lemma}
\newtheorem{Corollary}[Theorem]{Corollary}
\newtheorem{Remark}[Theorem]{Remark}

\makeatletter \@addtoreset{equation}{section}

\usepackage{color}

\def \E{\mathbb{E}}
\def \F{\mathbb{F}}

\def \P{\mathbb{P}}

\def \R{\mathbb{R}}

\def\Fc{{\cal F}}

\def\Mc{{\cal M}}

\def\Sc{{\cal S}}
\def\Tc{{\cal T}}

\def \0{\mathbf{0}}

\def\1{{\bf 1}}
\def \proof{{\noindent \bf Proof.\quad}}

\def \no{\noindent}
\def \ep{\hbox{ }\hfill{ $\Box$}}

 \title{Liquidation of an indivisible asset with independent investment}
 
 \author{Emilie {\sc Fabre} \footnote{CMAP, Ecole Polytechnique Paris, emilie.fabre@polytechnique.edu.}
 \and Guillaume {\sc Royer}\footnote{CMAP, Ecole Polytechnique Paris, guillaume.royer@polytechnique.edu.}
      \and Nizar {\sc Touzi}\footnote{CMAP, Ecole Polytechnique Paris, nizar.touzi@polytechnique.edu.
      Research supported by the Chair {\it Financial Risks} of the {\it Risk Foundation} sponsored by Soci\'et\'e
             G\'en\'erale, and
             the Chair {\it Finance and Sustainable Development} sponsored by EDF and Calyon. }}
             \date{\today}

\begin{document}

\maketitle

\begin{abstract}
We provide an extension of the explicit solution of a mixed optimal stopping -- optimal stochastic control problem introduced by Henderson and Hobson. The problem examines whether the optimal investment problem on a local martingale financial market is affected by the optimal liquidation of an independent indivisible asset. The indivisible asset process is defined by a homogeneous scalar stochastic differential equation, and the investor's preferences are defined by a general expected utility function. The value function is obtained in explicit form, and we prove the existence of an optimal stopping--investment strategy characterized as the limit of an explicit maximizing strategy. Our approach is based on the standard dynamic programming approach. 
\vspace{10mm}

 
\vspace{5mm}


\end{abstract}

\section{Introduction}

This paper considers a mixed optimal stopping/optimal control problem introduced by Henderson and Hobson \cite{hh} corresponding to the question of optimal liquidation of an indivisible asset in incomplete market, so that it is impossible to hedge it, while the agent is allowed to trade continuously in another financial asset. This question is motivated by problems on real option, see Dixit and Pindyck \cite{dp}, and Henderson \cite{h}.  

The framework of \cite{hh} is the following: an investor holds an indivisible asset, with price process defined as a geometric Brownian motion. In addition, a nonrisky asset, normalized to unity, and a financial asset are available for frictionless continuous-time trading. The risky asset price process is a local martingale with zero covariation with the indivisible asset process. The investor's preferences are defined by the expected power utility function. The objective of the risk averse investor is to choose optimally a stopping time for selling the indivisible asset, while optimally continuously trading on the financial market. 

In the absence of the indivisible asset, the problem reduces to a pure portfolio investment problem. Since the risky asset price process is a local martingale, it follows from the Jensen inequality that the optimal investment strategy of the risk averse investor consists in not trading the risky asset. Therefore, the main question raised by \cite{hh} is whether this optimal strategy is affected by the optimal liquidation problem of the independent indivisible asset. In the context of the power utility function, \cite{hh} shows that the answer to this question depends on the model parameters, and they provide the optimal stopping-investment strategies.

We observe that the last problem has been also studied in the case of correlated asset by Evans, Henderson and Hobson \cite{ehh}. We also refer to Henderson and Hobson for applications to American Options \cite{hh2}. 

In this paper we focus on the case where the financial asset is martingale and uncorrelated to the indivisible asset, corresponding to the setup of \cite{hh}. Our objective is to extend their results in two directions. First, the indivisible asset price process is defined by an arbitrary scalar homogeneous stochastic differential equation. Second, the investor's preferences are characterized by a general expected utility function. In contrast with \cite{hh}, we use the standard dynamic programming approach to stochastic control and optimal stopping to show that a lower bound is given by the limit of a sequence of functions defined by successive concavifications with respect to each variable. The resulting function is then the smallest majorant of the utility function which is partially concave in each of the variables. This construction of the lower bound induces a maximizing sequence of stopping times and portfolio strategies. This observation allows to prove that this lower bound indeed coincides with the value function. Finally, we prove that this maximizing sequence is weakly compact, and we deduce the existence of an optimal strategy. 

The construction of \cite{hh} depends strongly on the particular specification of the Black-Scholes dynamics for the price process of the illiquid asset, and the power utility function. Moreover, specific arguments are adopted to solve the problem in various cases. Our approach using dynamic programming equations as a guess allows to get an insight of the value in a general context. Moreover this characterization as an increasing limit of concave envelopes made computation of the value function more easy and tractable. In particular the interest of gambling while trying to liquidate an indivisible asset boils down to verify if the first concave function considered is equal to the value function. We illustrate as well our result by recomputing the results of \cite{hh} as this limit of concave envelopes in the Appendix.
 
The paper is organized as follows. The problem is formulated in Section \ref{sect:problemformulation}. The main results are stated in Section \ref{sect:mainresults}. In particular, in Subsection \ref{sect:puc}, we specialize the discussion to the original context of \cite{hh}, and we show that our general results cover their findings. The explicit derivation of the value function is reported in Section \ref{sect:valuefunction}. Finally, Section \ref{sect:existence} contains the proof of existence of an optimal stopping-investment strategy.

\section{Problem formulation}
\label{sect:problemformulation}

Let $B$ be a Brownian motion on a filtered probability space $ (\Omega, \mathcal{F}, \mathbb{F}:=\lbrace \mathcal{F}_t\rbrace_{t \geq 0}, \P)$. Throughout this paper, we consider an indivisible asset with price process $ Y^y $ defined by the stochastic differential equation:
$$ 
dY^y_t
= 
Y^y_t \big[\mu(Y^y_t)dt+\sigma(Y^y_t)dB_t\big], 
\;\;
Y_0^y=y>0
$$
where the coefficients $ \mu, \ \sigma \  : (0,\infty) \longrightarrow   \R$ are bounded, locally Lipschitz-continuous, and $\sigma>0$. In particular, this ensures the existence and uniqueness of a strong solution to the previous SDE.

The first objective of the investor is to decide about an optimal stopping time $\tau$ for the liquidation of the indivisible asset. We shall denote by $\Tc$ the collection of all finite $\F-$stopping times.

The financial market also allows for the continuous frictionless trading of a risky security whose price process is a local martingale orthogonal to $W$. Then assuming a zero interest rate (or, in other words, considering forward prices), the return from a self-financing portfolio strategy is a process $X$ in the set
\be\label{def:M}
\mathcal{M}^{\perp}(x)
:= 
\lbrace X \ \text{c\`adl\`ag martingale with} \ X_0=x, 
~\text{and} ~[X,B] =0 
\rbrace,
\ee
where $[X,B]$ denotes the quadratic covariation process between $ X$ and $B$. In the last admissibility set, the condition $[X,B]=0$ reflects that the indivisible asset cannot be partially hedged by the financial assets, while the martingale condition implies that, in the absence of the indivisible asset, the optimal investment in risky security of a risk-averse agent is zero. Following Hendersen and Hobson \cite{hh}, our objective is precisely to analyze the impact of the presence of the indivisible asset on this optimal no-trading strategy. 

Let $U:\R^+\longrightarrow \R \cup \lbrace-\infty\rbrace$ be a nondecreasing concave function, with $U>-\infty$ on $(0,\infty)$, be the utility function of a risk-averse investor. Our problem of interest is: 
\be\label{pb.V}
V(x,y)
&:=&
\sup_{(X,\tau)\in\Sc(x,y)} 
\E\big[U(X_{\tau}+Y^y_{\tau})\big], \ \ (x,y) \in D,
\ee
where $ D:= \lbrace \R \times (0,\infty) ; \ x+y \geq 0 \rbrace $,
$$
\Sc(x,y)
:=
\big\{(X,\tau)\in\Mc^\perp(x)\times\Tc: (X+ Y^y)_{.\wedge\tau} \geq 0 
                                                               ~\mbox{and}~
                                                               \lbrace U(X_{\tau\wedge\theta}+Y^y_{\tau \wedge\theta})^-\rbrace_{\theta\in\Tc }
                                                               ~\mbox{is UI}
 \big\},
$$
and UI is an abreviation for uniformly integrable.

We also introduce the corresponding no-trade problem:
\be\label{pb.m}
m(x,y)
&:=&
\sup_{\tau \in \mathcal{T}(x,y)} \E\big[U(x+Y^y_\tau)\big],
\ \ (x,y) \in D,
\ee
where $ \mathcal{T}(x,y):=\lbrace\tau\in\Tc:(x,\tau)\in\Sc(x,y) \rbrace$ and we abuse notation by identifying $x$ to the constant process equal to $x$.

While \eqref{pb.m} is a classical infinite horizon optimal stopping problem, we notice that \eqref{pb.V} is a mixed stochastic control--optimal stopping problem. We shall address both of them by means of the standard dynamic programming approach, see Fleming and Soner \cite{s} and Touzi \cite{lnt}. We observe that, similar to the last references, stochastic control and optimal stopping problems are studied separately in view of isolating the main arguments for the solution approach. However, it is well-known that mixed stochastic control--optimal stopping problems are easily addressed by an obvious superposition of the corresponding arguments.

\section{Main results}
\label{sect:mainresults}

\subsection{General utility function}

We first introduce a suitable change of variable, transforming the process $ Y^y$ into a local martingale. This is classically obtained by means of the scale function $ S$ of $ Y^y$ defined as a solution of:
$$ S'(y) y \mu(y) + \frac{1}{2} y^2 \sigma^2(y)S''(y)=0.$$
By additionally requiring that $S'(c)=1$ and $S(c)=0$, for some $c$ in the domain of the diffusion $Y$, this ordinary differential equation induces a uniquely defined continuous one-to-one function $S:(0,\infty)\longrightarrow dom(S)= \big( S(0),S(\infty) \big)$. We denote $R:=S^{-1}$ its continuous inverse. Then the process $ Z:=S(Y^y)$ is a local martingale satisfying the stochastic differential equation:
\b* 
dZ_t
=
\tilde{\sigma}(Z_t)dB_t ,
&\text{with}& 
\tilde{\sigma}(z)=R(z)S'(R(z))\sigma(R(z)).
\e*
From now on, we will work with the process $ Z$ instead of $ Y^y$. We define the corresponding domain 
\b* 
\bar{D}
&:=& 
\lbrace (x,z)\in \R \times dom(S) \ : \ x+R(z) \geq 0 \rbrace ,
\e* 
and we introduce the functions:
$$  
\bar{m}(x,z):=m(x,R(z)), 
\ \ 
\bar{V}(x,z):=V(x,R(z)) 
\ \ \text{and} \ \ 
\bar{U}(x,z):=U(x+R(z)),
\ \  (x,z)\in \bar{D}.
$$
Notice that $ \bar{U}$ is in general not concave w.r.t. $ z$ but still concave w.r.t. $ x$. We then introduce
$$ \bar{U}_1:=(\bar{U})^{\text{conc}_z},$$
where $ \text{conc}_z$ denotes the concave envelope w.r.t. $ z$.

We first give the characterization of  the no trade-problem, namely the value of $ m$:
\begin{Proposition}\label{m.calculation}
Assume that $ \bar{U}^1$ is locally bounded on $ int(\bar{D})$, then  $m(x,y)=\bar{U}^1(x,S(y))$ for all $ (x,y)\in\bar{D}$.
\end{Proposition}

\proof We organize the proof in three steps.\\
\textit{Step 1:} We first show that $ \bar{m} \le \bar{U}^1$. Let $ (x,z) \in \bar{D}$, $ \tau \in \Tc(x,R(z))$, $\theta_n$ a localizing sequence for the local martingale $ Z$, and define $ \tau_n=\tau \wedge \theta_n$. By the Jensen inequality, we have:
 \b*
 \E \left[ \bar{U}(x,Z_{\tau_n}) \right] 
 &\le& 
 \E \left[ \bar{U}^1(x,Z_{\tau_n}) \right] 
 \;\le\;  
 \bar{U}^1(x,\E[Z_{\tau_n}])
 \;=\;
 \bar{U}^1(x,z).
 \e*
Then, it follows from the Fatou Lemma that:
 \b*
 \liminf_{n\to\infty} \E \left[ \bar{U}(x,Z_{\tau_n})^+ \right] 
 &\ge& 
 \E \left[  \liminf_{n\to\infty} \bar{U}(x,Z_{\tau_n})^+ \right]
 \;=\; 
 \E \left[  \bar{U}(x,Z_{\tau})^+ \right]. 
 \e*
By the uniform integrability of the family $  \lbrace U(x+Y_{\tau \wedge\theta})^- ,\theta\in\Tc \rbrace$, we obtain:
 \b* 
 \lim_{n\to\infty} \E \left[ \bar{U}(x,Z_{\tau_n})^- \right]
 &=& 
 \E \left[ \bar{U}(x,Z_{\tau})^- \right].
 \e*
Then, $  \E \left[ \bar{U}(x,Z_{\tau}) \right] \le \bar{U}^1(x,z)$, and therefore $\bar m\le\bar U^1$, by the arbitrariness of $ \tau \in \Tc(x,R(z))$.
\\
\textit{Step 2:} For the second inequality we use the PDE characterization of the problem.  Let $ \bar m_*(x,z):=\underset{z'\rightarrow z , \ (x,z')\in \bar{D} }{\liminf} \bar{m}(x,z')$ be the lower semicontinuous envelop of the function $x\longmapsto \bar{m}(x,z)$. From Step 1, we have $\bar U\le \bar m\le\bar U^1$. Then, by the assumption that $\bar U^1$ is locally bounded, it follows that $\bar m_*$ is finite. By classical tools of stochastic control, we have that $ \bar{m}_*(x,\cdot)$ is a viscosity super-solution of:
$$ 
\min \lbrace u-\bar{U}(x,\cdot) , -u_{zz} \ \rbrace \geq 0,
$$
on $ int(\bar{D})$. Then it follows from Lemmas 6.9 and 6.23 in \cite{lnt} that $\bar{m}_*(x,z) \geq \bar{U}^1(x,z)$ for all $ (x,z)\in int(\bar{D})$. Combining with Step 1, we have thus proved that
\begin{align}\label{first prop.interm}
\bar{m} \le \bar{U}^1 \le \bar{m}_* \le \bar{m} \ \ \text{on} \ \  int(\bar{D}).
\end{align}
\textit{Step 3:} The property is also true on $ \partial \bar{D}$. Indeed for $ x\in \R$, denote $ z^x:= \inf \lbrace z\in dom(S) : x+R(z) \geq 0 \rbrace$. We then have $ \partial \bar{D} = \lbrace (x,z^x), \ x\in\R \ : z^x \in dom(S) \rbrace$. We now fix $ x$ such that $ (x,z^x)$ is in $\partial \bar{D}$.\\
Any element $ \tau \in \mathcal{T}(x,R(z^x))$ must verify $ Z^{z^x}_{\cdot \wedge \tau} \geq S(-x)$, with initial condition $ Z^{z^x}_0=S(-x)$. Now since $ Z^{z^x}$ is a local martingale and $ \tilde{\sigma}>0$, we deduce that $ \tau=0 $ a.s., so that $ \bar{m}(x,z^x)=\bar{U}(x,z^x)$.

Finally, since $ \bar{U}^1(x,\cdot)$ is the concave envelop of a continuous non-decreasing function defined on $ [z^x,S(\infty))$, it follows that $\bar{U}^1(x,z^x)=\bar{U}(x,z^x)$ at the left extreme point $z^x$ of the domain.
\ep

\vspace{5mm} 
We next return to our problem of interest $V$. Notice that $ \bar{U}^1$ is in general not concave in $x$, see the power utility example in Subsection \ref{sect:puc}. We remark also that the calculations performed in this context show that $ \bar{U}^1$ is not even continuous, in general, as illustrated by the case $ 1 <\gamma \le p$ of Proposition \ref{calcul.expl} in which we have $ \bar{U}^1$ locally bounded but discontinuous in the $ x$ variable (discontinuity at $ x=0$).

Since the risky asset price process is a local martingale, the value function is expected to be concave in $x$, because of the maximization over the trading strategies in the risky asset. We are then naturally lead to introduce a function $\bar U^2:=\left( \bar{U}^{1} \right)^{\text{conc}_x}$ as a further concavification of $\bar U^1$ with respect to the $x-$variable, which may again loose the concavity with respect to the $z-$variable. This leads naturally to the following sequence $ \left( \bar{U}^n \right)_n$:
$$  \bar{U}^0=\bar{U}, \ \ 
\bar{U}^{2n+1}=\left( \bar{U}^{2n} \right)^{\text{conc}_z} ,\ \
\bar{U}^{2n+2}=\left( \bar{U}^{2n+1} \right)^{\text{conc}_x}, \ \
n\ge 0 .
$$
The sequence  $ \left( \bar{U}^n \right)_n$ is clearly non decreasing, and then converges pointwise to a limit $ \bar{U}^{\infty}$ taking values in $ \R \cup \lbrace +\infty \rbrace $.
It is then easy to check that $ \bar{U}^{\infty}$ is the smallest dominant of  $\bar{U}$ which is partially concave in $x$, and partially concave in $ z$.

The first main result of the paper is the following:

\begin{Theorem} \label{solution}
Assume that the filtered probability space $(\Omega,\Fc,\F,\P)$ is sufficiently rich in the following sence:
\begin{itemize}
\item[(H1)] Either, there is a Brownian motion $W$ independent of $B$,
\item[(H2)] Or, there is a sequence $(\xi_n)_{n\ge 0}$ of independent uniformly distributed random variables which may be added to enrich the initial filtration.
\end{itemize}
Then, $V(x,y)=\bar{U}^\infty(x,S(y))$ for all $(x,y)\in D.$
In particular, $ V=m$ iff $ \bar{U}^\infty=\bar{U}^1$. Moreover, considering the restriction of $ \bar{U}^\infty$ on $ int( \bar{D})$, we obtain that if $ \bar{U}^\infty$ is locally bounded, then it is continuous. If $ \bar{U}^\infty$ is not locally bounded, then $ \bar{U}^\infty=+\infty$ on $ int(\bar{D})$.
\end{Theorem}

The proof that $ V(x,y) \le \bar{U}^\infty(x,S(y))$ is reported in section \ref{sect.upperbound}. The reverse inequality is proved under condition (H1) in section \ref{sect.lowerbound}, using the characterization by means of the dynamic programming equation, and under condition (H2) in section \ref{sect.maxsequence}, by building and explicit sequence of approximately optimal strategies .

Theorem \ref{solution} states that the value function is given by $ \bar{U}^\infty$. Explicit calculation of $ \bar{U}^\infty$ is possible in many cases as we will illustrate in Proposition \ref{calcul.expl} and the Appendix. In general, one may resort to numerical approximation techniques by exploiting the definition of $ \bar{U}^\infty$ as a limit of one-dimensional concave envelopes. 

The last part of Theorem \ref{solution} answers the economic relevant question raised by Hendersen and Hobson \cite{h}. Namely, at any starting point $(x,y)$, gambling has an interest for the investor if and only if $\bar{U}^\infty(x,S(y))=\bar{U}^1(x,S(y))$.

\begin{Remark}
Conditions (H1) and (H2) are necessary in order to allow for non-trivial orthogonal martingales. Indeed, if the probability space is not sufficiently rich, it may happen that $\mathcal{M}^{\perp}(x)$ is reduced to the constant process $x$, so that no gambling is possible and the problems $V$ and $m$ are identical. The last conclusion is not altered if one allows for randomized strategies along the standard relaxation technique in control theory. This shows that Conditions (H1) and (H2) are of different nature than the introduction of randomized strategies. 
\end{Remark}

We next focus on the existence and the characterization of a solution to the problem $ V$. We need to introduce the following assumption:

\begin{Assumption}\label{compact}
For all $ (x,z)\in int(\bar{D})$, there exists an open bounded subset $O:=O_{x,z}$ of $\bar{D}$, with $ (x,z)\in O$ and {\rm cl}$(O)\subset{\rm int}(\bar{D})$, such that $ \bar{U}=\bar{U}^\infty$ on $ \partial O$.
\end{Assumption}

Since $ \bar{U} \le \bar{U}^n \le \bar{U}^\infty$ for all $ n \geq 0$, this assumption implies that:
$$ \bar{U}^n=\bar{U} \ \text{on} \ \partial O \ \ \text{for all} \ n\geq 0.$$

\begin{Remark}
Assumption \ref{compact} implies that $ \bar{U}^\infty$ is locally bounded. This is a consequence of Lemma \ref{barUinftybdd} below.


\end{Remark}

The second main contribution of this paper is the following existence result.

\begin{Theorem}\label{opt.solution}
Let Assumption \ref{compact} hold true, and assume that the filtered probability space satisfies Condition (H2) of Theorem \ref{solution}. Then for all $ (x,y)\in D$:
$$ V(x,y)=\E[U(X^*_{\tau^*}+Y^y_{\tau^*})] \ \ \text{for some} \ \ \left(X^*,\tau^*\right) \in \Sc(x,y). $$
\end{Theorem}

This result is proved in section \ref{sect:existence}. The optimal strategy $ (X^*,\tau^*)$ will be characterized as the limit of an explicit sequence. Moreover if $ \bar{U}^\infty(x,z)=\bar{U}^n(x,z)$ for some $n$, then $ \left(X^*,\tau^*\right)$ is derived explicitly for the starting point $(x,z)$.

We shall argue in Remark \ref{rem-existencedoesnothold} below that Assumption \ref{compact} is necessary in the sense that we may find a situation where it is not satisfied and existence of optimal strategies fails.

\subsection{The power utility case}
\label{sect:puc}

In \cite{hh}, the indivisible asset $ Y^y $ is defined as a geometric Brownian motion:
$$ dY^y_t=Y^y_t (\mu dt + \sigma dB_t), \ \ Y^y_0=y>0$$
and the agent preferences are characterized by a power utility function with parameter $ p \in (0, \infty)$:
\begin{align}\label{def:U_p}
U_p(x)= \frac{x^{1-p}-1}{1-p}, \ \ p \neq 1, \ \ \text{and} \ \ U_1(x)=\ln(x).
\end{align}

Following \cite{hh}, we introduce the constants $ \gamma$ and $ \hat{\gamma}_p$ defined by:
$$ \gamma = \frac{2\mu}{\sigma^2} \ \ \text{and} \ \ \hat{\gamma}_p\in(0, p \wedge 1), \ (p-\hat{\gamma}_p)^p(p+1-\hat{\gamma}_p)-(2p-\hat{\gamma}_p)^p(1-\hat{\gamma}_p)=0,$$
where the existence and uniqueness of $ \hat{\gamma}_p$ follows from direct calculation.\\

\begin{Proposition}\label{calcul.expl}
Let $ U=U_p$ as defined in \eqref{def:U_p}. Then:
\\
{\rm (i)} for $ \gamma \le 0$, we have $ \bar{U}^\infty=\bar{U}^0 <\infty$,
\\
{\rm (ii)} for $ 0 < \gamma \le \hat{\gamma}_p$, we have $ \bar{U}^\infty=\bar{U}^1 <\infty$ and $ \bar{U}^0 \neq \bar{U}^1$,
\\
{\rm (iii)} for $ \hat{\gamma}_p < \gamma <1 \wedge p$, we have $ \bar{U}^\infty=\bar{U}^2 <\infty $ and $ \bar{U}^1 \neq \bar{U}^2$,
\\
{\rm (iv)} for $\gamma\ge p\wedge 1$, 
\\
{\rm $~$\quad(iv-a)} $p\le 1$, we have $\bar{U}^\infty=\bar{U}^2=+\infty$,
\\
{\rm $~$\quad(iv-b)} $p>1$, and $\gamma\le p$, we have $\bar{U}^\infty=\bar{U}^2<+\infty$,
\\
{\rm $~$\quad(iv-c)} $p>1$, and $\gamma> p$, we have $\bar{U}^\infty=\bar{U}^1<+\infty$.
\end{Proposition}

The proofs of these results are reported in the Appendix, and are obtained by explicit computation of the succession of concave envelopes.

\begin{Corollary}\label{corol:power utility}
Let $ U=U_p$ as defined in \eqref{def:U_p} and assume either (H1) or (H2). Then
\\
{\rm (i)} $ V=m$ if and only if $ \gamma \le \hat{\gamma}_p $ or $\gamma>p>1$,
\\
{\rm (ii)} for $ \gamma <p\wedge 1$, Assumption \ref{compact} holds true, so that under condition (H2) an optimal hedging-stopping strategy exists.
\end{Corollary}

\begin{Remark}
In the present power utility example, Proposition \ref{calcul.expl} states in particular that $\bar U^\infty$ equals either $U^0, U^1$, or $U^2$, whenever $\bar U^\infty<\infty$. Then, the optimal strategy is directly obtained from Lemma \ref{result.interm}, and there is no need to the limiting argument of Section \ref{sect:existence}. 
\end{Remark}

\begin{Remark}\label{rem-existencedoesnothold}
From our explicit calculations, we observe that Assumption \ref{compact} fails in cases (iv-b) and (iv-c) of Proposition \ref{calcul.expl}. Our explicit calculations in these cases show that $\bar U^\infty$ is asymptotic to $\bar U$ near infinity. Notice that this rules out the existence of an optimal strategy. Indeed, the optimal strategy in optimal stopping theory is given by the first hitting time of the obstacle by the value function. Therefore, if the value function is asymptotic to the obstacle, then such a hitting time may take the value $+\infty$, which is excluded by our definition of admissible stopping strategies.
\end{Remark}

The result of Corollary \ref{corol:power utility} is in line with the findings of \cite{hh}, and in fact complements with some missing cases there in. Loosely speaking, Corollary \ref{corol:power utility} states that when $ \gamma \le \hat{\gamma}_p$ or when $ \gamma>p>1$, the agent is indifferent to do fair investments on the market; the optimal strategy consists in keeping a constant wealth and solving an optimal stopping problem, i.e. $ m$. Instead, when $ \hat{\gamma}_p<\gamma \le p $, the agent can take advantage of a dynamic management strategy of its portfolio. 

\begin{Remark}
The methodology used in \cite{hh} is the following. 

- They construct a parametric family of stopping rules and admissible martingales by first fixing the portfolio value and waiting until the indivisible asset reaches a certain level, and then fixing the time and optimizing the jump of the portfolio value process. 

- For each element of this family, they evaluate the corresponding performance, and optimize over the parameter values. 

The rigorous proof follows from a verification argument. Our methodology relies on the standard dynamic programming approach to stochastic control, provides a better understanding of the value function $V$, and justifies the above construction of optimal strategies. Moreover, it shows that the result holds in a wider generality allowing for a larger class of utility functions and richer dynamics of the illiquid asset price process.
\end{Remark}

\section{Characterizing the value function}
\label{sect:valuefunction}

We first prove in section \ref{sect.upperbound} that $\bar V\le\bar U^\infty$. In section \ref{sect.lowerbound}, we prove the reverse inequality under Condition (H1) on the probability space. The corresponding result under Condition (H2) will be proved at the end of Subsection \ref{sect.maxsequence}. Notice that all of our subsequent proofs consider starting values $ (x,z)\in int(\bar{D})$. Indeed, recall from Step 3 of the proof of Proposition \ref{m.calculation} that $ \partial \bar{D} = \lbrace (x,z^x), \ x\in\R \ : z^x \in dom(S) \rbrace$, we obtain easily that $ \bar{U}^\infty =\bar{U}$ on $ \partial \bar{D} = \lbrace (x,z^x), \ x\in\R \ : z^x \in dom(S) \rbrace$, and that for any $ (x,z)\in \partial \bar{D}$, the corresponding set of admissible strategies $ \Sc(x,R(z)) $ is reduced to the constant martingale and stopping time equal to 0. 

\subsection{Upper bound}
\label{sect.upperbound}

\begin{Lemma} \label{barUinftybdd}
 $ \bar{U}^{\infty}$ is continuous on $ int(\bar{D})$ iff it is locally bounded. If $ \bar{U}^\infty$ is not locally bounded on $ int(\bar{D})$, then $ \bar{U}^\infty=+\infty$ on $ int(\bar{D})$. 
\end{Lemma}

\proof We first assume that $ \bar{U}^\infty$ locally bounded on $ int(\bar{D})$. Since $ \bar{U}^\infty$ is locally bounded, concave w.r.t. $ x $ and concave w.r.t. $ z$, we have that $ \bar{U}^\infty(x,\cdot)$ and $ \bar{U}^\infty(\cdot,z)$ are continuous on the interior of their domain, for all $ x$ and $ z$.

Now assume on the contrary that there exists $ \epsilon>0$, $ (x,z)  \in int(\bar{D}) $ and a sequence $ (x_n,z_n) \in int(\bar{D})$, $ (x_n,z_n)\underset{n \rightarrow +\infty}{\longrightarrow} (x,z)$ such that:
$$ \forall n \geq 0, \ \ |\bar{U}^\infty(x_n,z_n)-\bar{U}^\infty(x,z)| > \epsilon.$$
Without loss of generality, we assume that:
$$ \bar{U}^\infty(x_n,z_n) > \bar{U}^\infty(x,z)+\epsilon.$$
By continuity of $ \bar{U}^\infty(\cdot,z)$, we have for $ n$ large enough:
$$ \bar{U}^\infty(x_n,z_n)-\bar{U}^\infty(x_n,z)> \frac{\epsilon}{2}.$$
Without loss of generality, we assume $ z_n \geq z$ for all $ n \ge 0 $. We then define $ \tilde{z}^n=z-\sqrt{z_n-z}$, and observe from the concavity of $ \bar{U}^\infty(\cdot,z) $ that:
$$ \frac{ \bar{U}^\infty(x_n,z)-\bar{U}^\infty(x_n,\tilde{z}_n)}{z-\tilde{z}_n} \geq \frac{\bar{U}^\infty(x_n,z_n)-\bar{U}^\infty(x_n,z)}{z_n-z}>\frac{\epsilon}{2} \frac{1}{z_n-z}. $$
Then:
$$ \bar{U}^\infty(x_n,z)-\bar{U}^\infty(x_n,\tilde{z}_n)>\frac{\epsilon}{2} \frac{1}{\sqrt{z_n-z}}. $$
Since $ (x_n,\tilde{z}_n) \underset{n\rightarrow +\infty}{\longrightarrow} (x,z) $, this is a contradiction with the local boundedness of $ \bar{U}^\infty$.\\

We next assume that $ \bar{U}^\infty$ is not locally bounded on $ int(\bar{D})$, i.e. there exists $ (x,z)\in int(\bar{D})$ and $ (x_n,z_n)\rightarrow (x,z) $ such that $ \bar{U}(x_n,z_n)\rightarrow +\infty$. Let $ c>0$ be such that $ (x+c,z+c)\in int(\bar{D})$. Then it follows from the non-decrease of $\bar{U}$ in both variables $x$ and $z$ that
$$
\bar{U}^\infty(x+c,z+c)
\ge 
\bar{U}^\infty(x_n,z+c)
\ge
\bar{U}^\infty(x_n,z_n)
\longrightarrow\infty,~\mbox{as}~n\to\infty.
$$
Since  $ \bar{U}^\infty$ is partially concave w.r.t. to both variables, this implies that $ \bar{U}^\infty=\infty$ on $ \bar{D}$.
\ep 

\vspace{5mm} 
We now focus on the first inequality in Theorem \ref{solution}.

\begin{Lemma} \label{upbound}
$ \bar{V} \le \bar{U}^\infty$ on $ \bar{D}$.
\end{Lemma}

In order to prove Lemma \ref{upbound}, we use a regularization argument in the case $ \bar{U}^\infty$ locally bounded. By Lemma \ref{barUinftybdd}, $ \bar{U}^\infty$ is continuous on the interior of $ \bar{D}$. But in general, it is not twice differentiable in each variable. Therefore, we introduce for any $ \epsilon\in (0,1]$:
\begin{align}
 \bar{U}^n_\epsilon (x,z)  =\int_{\bar{D}} \bar{U}^n(\xi,\zeta)\rho_\epsilon(x-\xi,z-\zeta)d\xi d\zeta, \ (x,z)\in\bar{D}, \ \text{for all} \ n\in [0,\infty],
\end{align}
where for all $ u$ in $ \R^2$:
\b* 
\rho_\epsilon(u)=\epsilon^{-2} \rho(u/\epsilon) 
&\text{with}& 
\rho(u)=C e^{-1/(1-|u|^2)}{\bf{1}}_{|u|<1}, 
\e*
and $ C $ is chosen such that $ \int_{\R^2}\rho(u)du=\int_{B(0,1)}\rho(u)du=1$. Clearly, $\rho_\epsilon$ is $ C^\infty$, compactly supported, and $\rho_\epsilon$ converges pointwise to the Dirac mass at zero. We set by convention $ \bar{U}^n_0:=\bar{U}^n$ for $ \epsilon=0$.

We also intoduce for any $ \delta \geq 0$:
$$ \bar{U}^n_{\epsilon,\delta}(x,z):= \bar{U}^n_\epsilon(x+2\delta,z), \ (x+2\delta,z)\in \bar{D}, \  \text{for all} \ n\in [0,\infty].$$

\begin{Lemma} \label{U.properties}
$ \bar{U}_{\epsilon}^\infty \underset{\epsilon \rightarrow 0}{\longrightarrow} \bar{U}^\infty$ pointwise on  $ \bar{D}$, $\bar{U}_\epsilon^\infty \in C^\infty(\bar{D})$, $ \bar{U}_\epsilon^\infty \geq \bar{U}_\epsilon$ on $ \bar{D}$, and $ \bar{U}_{\epsilon,\delta}^\infty$ is partially concave in each variable, for all $ 0< \epsilon <\delta$.
\end{Lemma}

\proof
 The first three claims follow from classical properties of the convolution with respect to a non-negative kernel $ \rho_\epsilon$, together with the construction of $ \bar{U}^\infty$.\\
Let us prove the concavity of $ \bar{U}^\infty_{\epsilon,\delta}$ w.r.t. $ x$. The same proof holds for $ z$. For arbitrary $ 0<\epsilon<\delta$, we fix $ x$, $ x'$ and $ z$ such that $ (x,z)\in \bar{D}$ and $ (x',z)\in \bar{D}$. For $ \lambda \in [0,1]$, denote $ \hat{x}:=\lambda x+(1-\lambda)x'$. Then using the concavity of $ \bar{U}^\infty$ in $ x$:
\begin{align*}
\bar{U}^\infty_{\epsilon,\delta} (\hat{x},z) &= \int_{\R^2} \bar{U}^\infty(\lambda (x+2\delta+\xi)+(1-\lambda)(x'+2\delta+\xi),z+\zeta)\rho_\epsilon(\xi,\zeta) d\xi d \zeta \\
& \geq \int_{\R^2} \left( \lambda \bar{U}^\infty(x+2\delta+\xi,z+\zeta)+ (1-\lambda)\bar{U}^\infty(x'+2\delta+\xi,z+\zeta)  \right) \rho_\epsilon (\xi, \zeta) d \xi d \zeta \\
& = \lambda \bar{U}^\infty_{\epsilon,\delta} (x,z) + (1-\lambda) \bar{U}^\infty_{\epsilon,\delta} (x',z).
\end{align*}
\ep 

\vspace{5mm}

\no \textbf{Proof of Lemma \ref{upbound}} In the case $ \bar{U}^\infty$ not locally bounded, then by Lemma \ref{barUinftybdd}, we have $ \bar{U}^\infty=+\infty$ and the result is obvious.

Now assume that $ \bar{U}^\infty$ is locally bounded. We proceed in two steps.
\\
\textit{Step 1}. Let $ \left( \theta_n \right)_n$ be a localizing sequence for the local martingale $ Z$. We fix $ \delta >0$ and we consider $ 2\epsilon<\delta$. Let $ (X,\tau) \in \Sc(x,R(z)) $ and $ \tau_n=\tau \wedge \theta_n$. Clearely we have that $ (X,\tau_n)$ is in $ \Sc(x,R(z))$. Then by It\^o's formula for jump processes:
\begin{align*}
& \bar{U}^\infty_{\epsilon,\delta}(X_{t \wedge \tau},Z_{t \wedge \tau_n})-\bar{U}^\infty_{\epsilon,\delta}(x,z)= 
\\
& \int_0^{t \wedge \tau_n} \frac{1}{2} \partial_{xx}\bar{U}^\infty_{\epsilon,\delta}(X_u,Z_u)d[X,X]_u^c+\int_0^{t\wedge \tau_n} \frac{1}{2} \partial_{zz}\bar{U}^\infty_{\epsilon,\delta}(X_u,Z_u)\tilde{\sigma}^2(Z_u)du \\
& + \int_0^{t \wedge \tau_n} \partial_z \bar{U}^\infty_{\epsilon,\delta}(X_u,Z_u) \tilde{\sigma}(Z_u)dB_u+ \int_0^{t \wedge \tau_n} \partial_x \bar{U}^\infty_{\epsilon,\delta}(X_u,Z_u)dX_u 
\\
& + \underset{0<u\le t \wedge \tau_n}{\sum} \left( \bar{U}^\infty_{\epsilon,\delta}(X_u,Z_u)-\bar{U}^\infty_{\epsilon,\delta}(X_{u-},Z_u)-\partial_x \bar{U}^\infty_{\epsilon,\delta}(X_{u-},Z_u) \Delta X_u \right).
\nonumber
\end{align*}
Since $ \bar{U}^\infty_{\epsilon,\delta} $ is concave in $ x$ and in $ z$, then:
 \begin{equation}\label{barUinftyepsdelta}
 \bar{U}^\infty_{\epsilon,\delta}(X_{t \wedge \tau_n},Z_{t \wedge \tau_n})
 -\bar{U}^\infty_{\epsilon,\delta}(x,z) 
 \!\le \!\!
 \int_0^{t\wedge \tau_n}\!\!\! \partial_z \bar{U}^\infty_{\epsilon,\delta}(X_u,Z_u) \tilde{\sigma}(Z_u)dB_u 
 +\!\! \int_0^{t \wedge \tau_n} \!\!\!\partial_x \bar{U}^\infty_{\epsilon,\delta}(X_u,Z_u)dX_u.
 \end{equation}
We have for all $(\tilde{x},\tilde{z}) $:
\begin{align*}
\bar{U}_{\epsilon,\delta}(\tilde{x},\tilde{z}) &= \int_{\bar{B}((\tilde{x},\tilde{z}),\epsilon)}\bar{U}(\tilde{x}+2\delta-u,\tilde{z}-v) \rho_\epsilon(u,v)du dv \\
& \geq \int_{\bar{B}((\tilde{x},\tilde{z}),\epsilon)} U\left( \delta \right) \rho_\epsilon(u,v)du dv= U\left( \delta\right),
\end{align*}
where the last inequality follows from the fact that $ U$ is non decreasing and $ \tilde{x}+2\delta-u+R(\tilde{z}-v) \geq \delta$ on $ \bar{B}((\tilde{x},z),\epsilon)$. By Lemma \ref{U.properties}, this implies:
$$ \bar{U}^\infty_{\epsilon,\delta}(X_{t \wedge \tau_n},Z_{t \wedge \tau_n}) \geq U\left( \delta\right). $$

Since , $ | U \left(\delta\right)| < \infty $, the local martingale:
$$  \int_0^{t\wedge \tau_n} \partial_z \bar{U}^\infty_{\epsilon,\delta}(X_u,Z_u) \tilde{\sigma}(Z_u)dB_u + \int_0^{t \wedge \tau_n} \partial_x \bar{U}^\infty_{\epsilon,\delta}(X_u,Z_u)dX_u,
~~t \geq 0, 
$$
is bounded from below so it is a supermartingale. Then it follows from \eqref{barUinftyepsdelta} that:
$$ \E[\bar{U}_{\epsilon,\delta}^\infty(X_{t \wedge \tau_n },Z_{t\wedge \tau_n}] \le \bar{U}^\infty_{\epsilon,\delta}(x,z).$$

\no \textit{Step 2} Notice that $ \bar{U}^\infty_{\epsilon,\delta}(X_{t \wedge \tau_n},Z_{t \wedge \tau_n})$ is bounded from below by $ U(\frac{\delta}{2})$ Then, it follows from the pointwise convergence $ \bar{U}^\infty_{\epsilon,\delta}(x,z) \underset{\epsilon \rightarrow 0}{\longrightarrow} \bar{U}^\infty(x+2\delta,z)$, together with Fatou's Lemma, that:
\begin{align*}
\E\big[\bar{U}^\infty(X_{\tau}+2\delta,Z_\tau)\big] 
= 
\E \Big[\underset{\underset{\epsilon \rightarrow 0}{t,n\rightarrow \infty}}{\lim} 
            \bar{U}^\infty_{\epsilon,\delta}(X_{t \wedge \tau_n},Z_{t \wedge \tau_n}) 
    \Big]
&\le
\underset{\underset{\epsilon \rightarrow 0}{t,n \rightarrow \infty}}{\liminf} 
\E\big[\bar{U}^\infty_{\epsilon,\delta}(X_{t \wedge \tau_n},Z_{t \wedge \tau_n}) \big]\\
&\le \bar{U}^\infty(x+2\delta,z),
\end{align*}

By the arbitrariness of $(X,\tau)\in\Sc(x,R(z))$, this implies that $ \bar{V}(x,z)  \le \bar{U}^\infty(x+2\delta ,z)$, and therefore $ \bar{V}(x,z) \le \bar{U}^\infty(x,z)$, by the continuity of $ \bar{U}^\infty$ in the $ x$-variable.
\ep

\subsection{Lower bound for the value function under (H1)}
\label{sect.lowerbound}

Under Assumption (H1) on the filtration, it follows that $ \mathcal{M}^\perp$ is non-trivial, and contains the set:
$$ 
\mathcal{M}^W(x):= \lbrace X \ C^0\text{-mart}: \ X_t= x +\int_0^t \phi_s dW_s \ \text{for some} \ \phi\in \mathbb{H}^2_{\text{loc}} \rbrace.$$ 
In this subsection, we use the PDE characterization of the problem to obtain the lower bound for the value function. In order to use the classical tools of stochastic control and viscosity solutions we introduce the following simplified problem $ V^0$:
$$ V^0(x,y):= \underset{(X,\tau)\in \Sc^W(x,y)}{\sup} \E[U(X_{\tau}+Y_{\tau}^{y})],$$
where $ \Sc^W(x,y):= \big\{ (X,\tau)\in \Sc(x,y): X\in \mathcal{M}^W(x) \big\}$.\\

Since $ \mathcal{M}^W (x) \subset \mathcal{M}^\perp (x) $, we have
$$ V^0(x,y) \le V(x,y).$$
The reason for introducing the problem $V^0$ is that it has the standard form of a (singular) stochastic control problem which may be addressed by the classical dynamic programming approach. We recall the definition of the lower semi-continuous envelope:
 \b*  
 V^0_*(x,y)
 &:=&
 \liminf_{(x',y')\to (x,y)} V^0(x,y) , 
 ~~(x,y)\in D. 
 \e*

By Lemma \ref{upbound}, we have $ U(x+y)\le V^0(x,y)\le V(x,y) \le \bar{U}^\infty(x,R(y))$. Then, if $ \bar{U}^\infty$ is locally bounded, so is $V$, and it follows that $ V^0_*$ is finite.

We now derive the dynamic programming equation, which will provide us with the lower bound. For notation simplicity, we shall use subscripts to indicate partial derivatives with respect to the corresponding variables.

\begin{Proposition}\label{supersol}
Assume that $ \bar{U}^\infty$ is locally bounded, then $ \bar{V}^0_*$ is a viscosity supersolution of:
$$ \min \lbrace -v_{zz},-v_{xx},v-\bar{U}\rbrace=0 \ \ \text{on} \ int(\bar{D}). $$
In particular $ \bar{V}^0_*$ is partially concave w.r.t $ x$ and $ z$.
\end{Proposition}

\proof We first show that $ V^0_*$ is a viscosity supersolution of:
\begin{align} \label{edpV0}
\min \lbrace -\frac{1}{2}y^2 \sigma(y)^2 v_{yy}(x,y)-y \mu(y)v_y(x,y);-v_{xx}(x,y);v-U(x+y) \rbrace=0
\end{align}
on $ D$. First, since immediate selling of the illiquid asset is legitimate, we see that $V^0(x,y)\ge U(x+y)$, and therefore $V^0_*(x,y)\ge U(x+y)$, by the continuity of $U$.

We next continue by using the first part of the weak dynamic programming principle of Theorem 4.1 in \cite{bt}:
 $$ 
 V^0(x,y) 
 \geq 
 \underset{(X,\tau)\in \Sc^W(x,y)}{\sup} 
 \E \left[ V^0_*(X_\theta,Y^y_\theta){\bf{1}}_{\theta \le \tau}
            +U(X_\theta+Y^y_\theta){\bf{1}}_{\theta > \tau}
      \right]  
  \ \text{for all} \ \theta \ \text{stopping time}.
 $$
In order to justify the viscosity supersolution property, we consider an interior point $(x_0,y_0)$ and a test function $ \phi\in \mathcal{C}^{2,2}(\R)$ such that:
 $$ 
 \min(V^0_*-\phi)=(V^0_*-\phi)(x_0,y_0)=0.
 $$ 
Let $ (x_n,y_n)_{n\geq0}$ be a sequence such that $ (x_n,y_n,V^0(x_n,y_n))\rightarrow (x_0,y_0,V^0_*(x_0,y_0))$ as $ n\rightarrow\infty$.

For fixed $ \alpha \in \R$, define  $(X^n,Y^n) :=(x_n+\alpha W_{\cdot \wedge \theta_n},Y^{y_n}_{\cdot \wedge \theta_n}) $,
where $c>0$ is a constant, and:
 $$ \theta_n 
 := 
 h_n\wedge\inf \big\{ t \geq 0 :~ |X^n_t-x_n|+|Y^n_t-y_n|\ge c\big\},
 $$
with 
 \b*
 h_n:=\sqrt{|\beta_n|}{\bf{1}}_{\beta_n \ne 0}+ \frac{1}{n} {\bf{1}}_{\beta_n=0} 
 &\mbox{where}& 
 \beta_n:=V^0(x_n,y_n)-\phi(x_n,y_n)\to 0.
 \e*
Since $(x_0,y_0)$ is in the interior of the domain, notice that we may choose the constant $c>0$ sufficiently small so as to ensure that $(X^n,\theta_n)\in \Sc^W(x_n,y_n)$.\\
By the dynamic programming principle together with It\^o's formula, it follows that:
 \begin{align*}
 V^0(x_n,y_n) 
 &=
 \beta_n+\phi(x_n,y_n) \ge \E[\phi(X^n_{\theta_n},Y^n_{\theta_n})] 
 \\
 &= 
 \phi(x_n,y_n)
 +\E \Big[\int_0^{\theta_n}\Big( y\mu \phi_y + \frac{1}{2} y^2 \sigma^2 \phi_{yy}
                                                + \frac{1}{2} \alpha^2 \phi_{xx} \Big) 
                                          \big(X^n_u, Y^n_u\big) du 
       \Big].
\end{align*}
This leads to:
\b*
\beta_n 
&\ge& 
\E\Big[ \int_0^{\theta_n}\Big( y\mu \phi_y + \frac{1}{2} y^2 \sigma^2 \phi_{yy}+ \frac{1}{2} \alpha^2 \phi_{xx} \Big) \left(X^n_u, Y^n_u \right) du\Big]
\e*
Since $ \mu$ and $ \sigma$ are locally Lipschitz continuous and have linear growth, one can show the following standard estimate for all $ h>0$:
$$ \E \left[ \underset{t \le s \le t+h}{\sup}|Y^{y_n}_s-y_n|^2 \right] \le C h^2(1+|y_n|^2).$$
This leads to $ (X^n,Y^n) \underset{n \rightarrow \infty}{\longrightarrow}(x_0+\alpha W,Y^{y_0}) $ $ \P$-a.s. Moreover by definition of $ \theta_n$, the following quantity 
\begin{align*}
 \frac{1}{h_n} \int_0^{\theta_n}\left( y\mu \phi_y + \frac{1}{2} y^2 \sigma^2 \phi_{yy}+ \frac{1}{2} \alpha \phi_{xx} \right) \left(X^n_u, Y^n_u \right) du
\end{align*}
is bounded, uniformly in $ n$. Therefore, by the mean value and the dominated convergence theorem,
\b*
0
&\ge&
\frac{1}{2} y_0^2 \sigma^2(y_0)\phi_{yy}(x_0,y_0)+y^0\mu(y_0)\phi_y(x_0,y_0)+\frac{1}{2}\alpha^2 \phi_{xx}(x_0,y_0).
\e*
By the arbitrariness of $\alpha\in\R$, this implies that $ -\phi_{xx}(x_0,y_0) \le 0$. Hence, $ V^0_*$ is a viscosity supersolution on $ D$ of:
$$ \min \lbrace -\frac{1}{2}y^2 \sigma^2(y)v_{yy}-y\mu(y)v_y ; \ -v_{xx} ; \ v(x,y)-U(x+y) \rbrace =0.$$

Finally, the supersolution stated in the proposition is a direct consequence of the first step and the change of variable in the theory of viscosity solutions, see e.g. \cite{s}. The partial concavity property follows from Lemmas 6.9 and 6.23 in \cite{lnt}. 
\ep

\begin{Corollary}
Assume $ \bar{U}^\infty$ is locally bounded. Then for all $ (x,y) \in D $, we have:
$$ V(x,y) \geq \bar{U}^\infty (x,S(y)).$$
\end{Corollary}

\proof
We already know that $ V(x,y) \geq V^0(x,y) \geq \bar{V}_*^0(x,S(y)) $. On the other hand, since $ \bar{V}_*^0 $ is partially concave w.r.t. $ x$ and w.r.t. $ z$, and is a majorant of $ \bar{U}$, it follows that $ \bar{V}_*^0 $ is a majorant of $ \bar{U}^\infty$. This completes the proof.
\ep

\section{Optimal strategy}
\label{sect:existence}

We now derive an optimal strategy under Assumption \ref{compact} together with Condition (H2) of Theorem \ref{solution}. This will allow also to recover the case $ \bar{U}^\infty=+\infty$ since the construction is robust, whenever the concave envelopes are not finite.

\subsection{Construction of a maximizing sequence under (H2)}
\label{sect.maxsequence}

We fix $ (X_0,Z_0)=(x,z)\in int(\bar{D})$ and we consider $ O :=O_{x,z}$ the open set defined in Assumption \ref{compact}. We define the following sequence of stopping times $ (\tau^n)_{n \geq 0}$:\\

Since $ \bar{U}^1$ is the concavification of $ \bar{U}$ with respect to the $ z$-variable, we introduce the stopping time with frozen x-variable:
$$ \tau^0_1=\inf\lbrace t \geq 0 : \ \bar{U}^1(X_0,Z_t)=\bar{U}^0(X_{0},Z_t) \rbrace,$$
At time $ \tau_1^0$, $ Z_{\tau_1^0}$ takes values in $ \lbrace z_1,z_2 \rbrace $ where $ z_1 = \sup \lbrace z\le Z_0: \ \bar{U}^1(X_0,z)=\bar{U}(X_0,z) \rbrace $ and $ z_2 =\inf \lbrace z\geq Z_0: \ \bar{U}^1(X_0,z)=\bar{U}(X_0,z) \rbrace$. Notice that $z_1$ and $z_2$ anre finite, since $ (X_0,z_1)$ and $ (X_0,z_2)$ are both in ${{\rm cl}(O)}$. We then define:
 \b* 
 X_t
 :=
 X_0\1_{\{t < \tau^0_1\}}
 +\eta(X_0,Z_{\tau^0_1})\1_{\{t \ge \tau^0_1\}},
 \e*
where $ \E \left[ \eta(X_0,Z_{\tau^0_1})| \mathcal{F}_{\tau^0_1-} \right]=X_0 $ and:
 \b* 
 \P \big[ \eta(X_0,Z_{\tau^0_1})=a(X_0,Z_{\tau^0_1}) |(X_0,Z_{\tau^0_1})  \big]
 &=&
 p(X_0,Z_{\tau^0_1}),
 \\
 \P \big[ \eta(X_0,Z_{\tau^0_1})=b(X_0,Z_{\tau^0_1})|(X_0,Z_{\tau^0_1})  \big]
 &=&
 1-p(X_0,Z_{\tau^0_1}),
 \e*
with:
\begin{align*}
& d(v):= \lbrace x \in \R : (x,v)\in \bar{D} \rbrace , \\
& a(u,v):=\inf \lbrace \alpha \in d(v), \ \alpha \geq u : \bar{U}^{2}(\alpha,v)=\bar{U}^{1}(\alpha,v) \rbrace , \\
& b(u,v):=\sup \lbrace \alpha \in d(v), \ \alpha \le u : \bar{U}^{2}(\alpha,v)=\bar{U}^{1}(\alpha,v) \rbrace ,
\end{align*}
and $ p(u,v)$ is defined by:
$$ u=p(u,v)a(u,v)+(1-p(u,v))b(u,v).$$
Similarly, we define a sequence of stopping times $(\tau_i^n)_{0\le i\le n+1}$ by $ \tau^n_0=0$, and:
 $$ 
 \tau^n_{i}:= 
 \inf\big\{ t \geq \tau^n_{i-1}:~
               \bar{U}^{(2(n-i+1)+1}(X^n_{\tau^n_{i-1}},Z_{t})
               =\bar{U}^{2(n-i+1)}(X^n_{\tau^n_{i-1}},Z_{t}) 
      \big\},
 ~1\le i\le n+1,
 $$
where the martingale $ X^n$ is constructed as follows. Let:
\begin{align*}
& a^n_i(u,v):=\inf \big\{ \alpha \in d(v), \ \alpha \geq u : \bar{U}^{2(n-i+1)}(\alpha,v)=\bar{U}^{2(n-i+1)-1}(\alpha,v) \big\} , \\
& b^n_i(u,v):=\sup \big\{ \alpha \in d(v), \ \alpha \le u : \bar{U}^{2(n-i+1)}(\alpha,v)=\bar{U}^{2(n-i+1)-1}(\alpha,v) \big\} .
\end{align*}
By Assumption \ref{compact},  $ (a^n(u,v),v) $ and $ (b^n(u,v),v)$ are in ${{\rm cl}(O)}$ and $ \bar{U}^{2n-i+1}(\cdot,v)$ is linear on $ [a^n_i(u,v),b^n_i(u,v)]$. We then define $ p^n_i(u,v)\in[0,1]$ by:
$$ 
u=p^n_i(u,v)a^n_i(u,v)+(1-p^n_i(u,v))b^n_i(u,v),
$$
so that:
$$ 
\bar{U}^{2(n-i+1)}(u,v) =p^n_i(u,v)\bar{U}^{2(n-i+1)-1}(a^n_i(u,v),v)+(1-p^n_i(u,v))\bar{U}^{2(n-i+1)-1}(b^n_i(u,v),v).$$
With these notations, we define the process $ X^n$:\\
$$ 
X^n_t=X^n_0{\bf{1}}_{[0,\tau^n_1)}(t)+\sum_{i=1}^{n-1} \eta^n_i(X^n_{\tau^n_{i-1}},Z_{\tau^n_i}){\bf{1}}_{[\tau^n_i,\tau^n_{i+1})}(t)+\eta^n_n(X^n_{\tau^n_{n-1}},Z_{\tau^n_n}){\bf{1}}_{[\tau^n_n,\infty)}(t), 
$$
where each r.v. $ \eta^n_i(X^n_{\tau^n_{i-1}},Z_{\tau^n_i})$ is independant of $ \mathcal{F}_{\tau^n_i}$ and has distribution:
\begin{align*}
 &\P\left[ \eta^n_i(X^n_{\tau^n_{i-1}},Z_{\tau^n_i})=a^n_i(X^n_{\tau^n_{i-1}},Z_{\tau^n_i}) |\mathcal{F}_{\tau^n_{i}-} \right]=p^n_i(X^n_{\tau^n_{i-1}},Z_{\tau^n_i}) , \\
& \P\left[ \eta^n_i(X^n_{\tau^n_{i-1}},Z_{\tau^n_i})=b^n_i(X^n_{\tau^n_{i-1}},Z_{\tau^n_i})|\mathcal{F}_{\tau^n_{i}-}\right] =1-p^n_i(X^n_{\tau^n_{i-1}},Z_{\tau^n_i}).
\end{align*}
The existence of such r.v. $ \{\eta_i^n,i\le n\}_n$ is guaranteed by Assumption (H2).

\begin{Remark}
There is no issue of measurability of $ p^n_i, a^n_i $ and $ b^n_i$ as they are only involved in a finite number of values at each step.
\end{Remark}

\begin{Lemma}
Under assumption \ref{compact}, $(X^n,\tau^n_{n+1}) \in \Sc(x,y)$ for all $ n \geq 1$.
\end{Lemma}

\proof  $ [X^n,Z]=0$ follows from the fact that $ X $ is a pure jump process and $ Z$ is continuous. We also see that $ (X^n,Z)$ takes values in the compact cl$(O)$ given by assumption \ref{compact}, so $ \tau^n_{n+1} \in \mathcal{T}$ and the process $X^n+R(Z)$ is non negative. We now prove the martingale property. For all $ i \in \lbrace 1,...,n\rbrace$:
\begin{itemize}
\item $ t \in (\tau^n_i,\tau^n_{i+1}) \Rightarrow \E[X_t^n|\mathcal{F}_{t-}]=X^n_{t-}, $
\item If $ t=\tau^n_i$, then:
\begin{align*}
\E[X^n_t|\mathcal{F}_{t-}] &= \E[\eta^n_i(X^n_{\tau^n_{i-1}},Z_{\tau^n_i})|\mathcal{F}_{t-}] \\
&= a^n_i(X^n_{\tau^n_{i-1}},Z_{\tau^n_i})\E[{\bf{1}}_{\eta^n_i=a^n_i}|\mathcal{F}_{t-}] + b^n_i(X^n_{\tau^n_{i-1}},Z_{\tau^n_i}) \E[1-{\bf{1}}_{\eta^n_i=a^n_i}|\mathcal{F}_{t-}]\\
&=X^n_{\tau^n_{i-1}} =X^n_{t-}.
\end{align*}
\end{itemize}
\ep

\vspace{5mm}

The crucial property of the sequence $(X^n,\tau^n_{n+1})_n$ is the following.

\begin{Lemma} \label{result.interm}
For all $ n \geq 0 $, we have:
\begin{align}
\E[\bar{U}(X^n_{\tau^n_{n+1}},Z_{\tau^n_{n+1}})]=\bar{U}^{2n+1}(x,z). \label{part.sol}
\end{align}
\end{Lemma}

\proof We organize the proof in three steps.
\\
\textit{Step1:} We first show that for all $ i\in \lbrace 1,...,n+1\rbrace$, we have:
\begin{align}
\E\left[ \bar{U}^{2(n-i+1)-1} \left(X^n_{\tau^n_i},Z_{\tau^n_i} \right) \right]=\E\left[ \bar{U}^{2(n-i+1)} \left(X^n_{\tau^n_{i-1}},Z_{\tau^n_i} \right) \right]. \label{interm1}
\end{align}
Indeed:
\begin{align*}
\E\left[ \bar{U}^{2(n-i+1)-1} \left(X^n_{\tau^n_i},Z_{\tau^n_i} \right) \right] &= \E [ \bar{U}^{2(n-i+1)-1} \left( a^n_i(X^n_{\tau^n_{i-1}},Z_{\tau^n_i} ),Z_{\tau^n_i} \right) \E \left[{\bf{1}}_{\eta_i^n=a_i^n}| X^n_{\tau^n_{i-1}},Z_{\tau^n_i}\right] \\
& + \bar{U}^{2(n-i+1)-1} \left( b^n_i(X^n_{\tau^n_{i-1}},Z_{\tau^n_i} ),Z_{\tau^n_i} \right) \E \left[{\bf{1}}_{\eta_i^n=b_i^n}| X^n_{\tau^n_{i-1}},Z_{\tau^n_i}\right] ] \\
&= \E [ \bar{U}^{2(n-i+1)-1} \left( a^n_i(X^n_{\tau^n_{i-1}},Z_{\tau^n_i} ),Z_{\tau^n_i} \right) p^n_i(X^n_{\tau^n_{i-1}},Z_{\tau^n_i})\\
&+ \bar{U}^{2(n-i+1)-1} \left( b^n_i(X^n_{\tau^n_{i-1}},Z_{\tau^n_i} ),Z_{\tau^n_i} \right) (1-p^n_i(X^n_{\tau^n_{i-1}},Z_{\tau^n_i})).
\end{align*}
Then by definition of the random variables $ a^n_i(X^n_{\tau^n_{i-1}},Z_{\tau^n_i} ) $ and $ b^n_i(X^n_{\tau^n_{i-1}},Z_{\tau^n_i} )$, and the linearity of $ \bar{U}^{2(n-i+1)}(\cdot,Z_{\tau^n_i})$ on $ \left[ b^n_i(X^n_{\tau^n_{i-1}},Z_{\tau^n_i} ),a^n_i(X^n_{\tau^n_{i-1}},Z_{\tau^n_i} )\right]$, we have:
\begin{align*}
\E\left[ \bar{U}^{2(n-i+1)-1} (X^n_{\tau^n_i},Z_{\tau^n_i} ) \right] &=\E \left[ \bar{U}^{2(n-i+1)} \left( a^n_i(X^n_{\tau^n_{i-1}},Z_{\tau^n_i} ),Z_{\tau^n_i} \right) p^n_i(X^n_{\tau^n_{i-1}},Z_{\tau^n_i}) \right. \\
& \left. + \bar{U}^{2(n-i+1)} \left( b^n_i(X^n_{\tau^n_{i-1}},Z_{\tau^n_i} ),Z_{\tau^n_i} \right) (1-p^n_i(X^n_{\tau^n_{i-1}},Z_{\tau^n_i})) \right] \\
&=\E\left[ \bar{U}^{2(n-i+1)} (X^n_{\tau^n_{i-1}},Z_{\tau^n_i} ) \right].
\end{align*}
\textit{Step 2:} We next show that:
\begin{align}
\E\left[ \bar{U}^{2(n-i+1)} (X^n_{\tau^n_{i-1}},Z_{\tau^n_i} ) \right]=\E\left[ \bar{U}^{2(n-i+1)+1} (X^n_{\tau^n_{i-1}},Z_{\tau^n_{i-1}} ) \right]. \label{interm2}
\end{align}
We emphasize here that the process $ X^n$ takes its values in a finite set. Then the fact that $  \sigma>0$ and continuous ensures that $ |\tilde{\sigma}|>c>0$ on $ \text{proj}_z\big({\rm cl}(O)\big)$ and then if follows that for all $ i$, $ \tau^n_i <\infty$  and that $ \E[X^n_{\tau^n_i}|X^n_{\tau^n_{i-1}}]=X_{\tau^n_{i-1}}$.\\
 Then we know that $ \bar{U}^{2(n-i+1)+1} \left(X^n_{\tau^n_{i-1}},z \right) $ is linear on $ H_i^n$ where:
$$ H^n_i:= \big\{ z>0 \ : \ \bar{U}^{2(n-i+1)+1} (X^n_{\tau^n_{i-1}},z )>\bar{U}^{2(n-i+1)} (X^n_{\tau^n_{i-1}},z ) \big\}.$$
We can now conclude, by definition of $ \tau^n_i$ that:
\begin{align*}
 \E\left[ \bar{U}^{2(n-i+1)} (X^n_{\tau^n_{i-1}},Z_{\tau^n_i}) \right] &=\E\left[ \bar{U}^{2(n-i+1)+1} (X^n_{\tau^n_{i-1}},Z_{\tau^n_{i}} ) \right] \\
 &=\E\left[ \bar{U}^{2(n-i+1)+1} (X^n_{\tau^n_{i-1}},Z_{\tau^n_{i-1}} ) \right].
\end{align*}
\\
\textit{Step 3:} we now prove \eqref{part.sol}:
Using \eqref{interm1} and \eqref{interm2} we have:
\begin{align*}
\bar{U}^{2n+1}(x,z) = &\sum_{i=1}^n \E\left[ \bar{U}^{2(n-i+1)}(X^n_{\tau^n_{i-1}},Z_{\tau^n_{i}})-  \bar{U}^{2(n-i+1)-1}(X^n_{\tau^n_{i}},Z_{\tau^n_{i}})\right] \\
&+ \sum_{i=0}^n \E \left[ \bar{U}^{2(n-i+1)-1}(X^n_{\tau^n_{i-1}},Z_{\tau^n_{i}})-  \bar{U}^{2(n-i+1)-2}(X^n_{\tau^n_{i}},Z_{\tau^n_{i+1}})\right] \\
&+ \E\left[ \bar{U}^0 (X^n_{\tau^n_n},Z_{\tau^n_{n+1}})\right] \\
=& \E\left[ \bar{U}^0 (X^n_{\tau^n_n},Z_{\tau^n_{n+1}})\right].
\end{align*}
By construction, we have $\tau^n_{n+1} \geq \tau^n_n $ so we have $ X^n_{\tau^n_{n+1}}=X^n_{\tau^n_n}$ and then:
$$ \bar{U}^{2n+1}(x,z)=\E\left[ \bar{U}^0 (X^n_{\tau^n_{n+1}},Z_{\tau^n_{n+1}})\right]. $$
\ep

\vspace{5mm}

\no {\bf Proof of Theorem \ref{solution} under (H2)}
By Lemma \ref{upbound}, $ \bar{V} \le \bar{U}^\infty$. Then, since the sequence $ \left( \bar{U}^n \right)_n $ converges towards $\bar{U}^\infty$, it follows immediately from Lemma \ref{result.interm} that $(X^n,\tau^n_{n+1})_n$ is a maximizing sequence of strategies.
\ep

\begin{Remark}
Notice that Assumption \ref{compact} and the local boundedness condition of $ \bar{U}^\infty$ are not necessary to obtain a maximizing sequence. Indeed we have that the concave envelope $ f^{\text{conc}}$ of a function $ f$ defined on an interval $ I \subset \R$ is given by:
 $$
 \underset{\underset{y_1,y_2\in I}{y_1 \le y \le y_2}}{\sup} 
 \big\{ \lambda(y_1,y_2) f(y_1)+(1-\lambda(y_1,y_2))f(y_2)
 \big\}, \ \text{with} \  \lambda(y_1,y_2)=\frac{y_2-y}{y_2-y_1},
 $$
with the convention $ \lambda(y,\cdot)=1$ and $ \lambda(\cdot,y)=0$. So we could have considered $ \epsilon$-optimal sequences of coefficients $ a^n_i$ and $ b^n_i$ rather than optimal ones, which may not exist in the general case, and the proof holds. However the present construction is crucial for the existence result of the subsequent section.
\end{Remark}

\subsection{Existence of an optimal strategy}

\textbf{Proof of Theorem \ref{opt.solution}} Let $ ( X^n_{\tau^n_{n+1}},Z_{\tau^n_{n+1}})_{n \geq 0}$ be the sequence defined in Lemma \ref{result.interm}.
These pairs of random variables take values in the compact subset ${{\rm cl}(O)}$. We then define $ \mu_n$ the law of $ ( X^n_{\tau^n_{n+1}},Z_{\tau^n_{n+1}})$. This is a sequence of probability distributions with support in the compact subset ${{\rm cl}(O)}$. Then $ (\mu_n)$ is tight, and by the Prokhorov theorem we may find a subsequence, still renamed $ (\mu_n)$, which converges to some probability distribution $ \mu$ with support in ${{\rm cl}(O)}$.
\\
\textit{Step 1:} We first prove that $ \int_{{\rm cl}(O)} \bar{U}(\xi,\zeta)d\mu(\xi,\zeta)=\bar{U}^\infty(x,z)$.\\
Indeed, we have that $ \bar{U}$ is continuous on $ \bar{D}$ and ${{\rm cl}(O)}$ is a compact of $ \bar{D}$, So by Lemma \ref{result.interm} together with the weak convergence property,  we obtain:
$$
\bar{U}^\infty(x,z)= \underset{n\rightarrow \infty}{\lim} \bar{U}^n(x,z)
= 
\underset{n\rightarrow \infty}{\lim} \int_{{\rm cl}(O)} \bar{U}(\xi,\zeta) d\mu^n(\xi,\zeta) 
= 
\int_{{\rm cl}(O)} \bar{U}(\xi,\zeta) d\mu(\xi,\zeta) .
$$
\textit{Step 2:} We next introduce a pair $ (X^*,\tau^*)$ such that $ (X^*_{\tau^*},Z_{\tau^*})\sim \mu$.

First, we consider $ \tau^*$ a $ \left( \sigma(B_{0 \le s \le t }) \right)_{t\geq 0} $-stopping time such that $ Z_{\tau^*} \sim \mu_z$, where $ \mu_z(A):= \int_{\R \times A}\mu(dx,dz)$ is the $ z$-marginal law of $ \mu$. Such a stopping time exists because $ \mu_z$ is compactly supported and $\tilde{\sigma} \geq c >0 $ on ${{\rm cl}(O)} $ for some $ c>0$, thanks to  the assumption that $ \sigma >0$. Moreover we consider this stopping time such that $ \tau^*$ is smaller than the exit time of the support of $ \mu_z$. This result is proved in \cite{h}, section 4.3, or in Monroe \cite{m}.

We now consider  $ f : [0,1]^2 \rightarrow K $ a Borel function such that the pushforward measure of the lesbegue measure on $ [0,1]^2$ by $ f$ is $ \mu$ and $ f(x,y)=(f_1(x,y),f_2(y))$. The existence of this function corresponds to the existence of the conditional probability distribution.

We denote $ F_{\mu_z}$ the cumulative distribution function of $ \mu_z$. $ \zeta$ denotes a uniform random variable independent of $ B$ and we implicitly assume that the filtration $ \mathbb{F}$ is rich enough to support that $ \zeta$ is $ \mathcal{F}_{\tau^*}$-measurable and independant of $ \mathcal{F}_{\tau^*-}$. In particular, $ \zeta$ is independent of $ \sigma(B_{0 \le s \le \tau^* })$.

The candidate process $ X^*$ is then:
 $$ 
 X^*_t
 :=
  f_1(\zeta,F_{\mu_z}(Z_{\tau^*})) {\bf{1}}_{t \geq \tau^*},
  ~~t\ge 0.
  $$
Then we clearely have that $ (X^*_{\tau^*},Z_{\tau^*}) \sim \mu$.\\

\textit{Step 3:} It remains to prove that the pair $ (X^*,\tau^*)$ is in $ \mathcal{S}(x,R(z))$. We first show that $ X^*$ is a martingale in $ \mathcal{M}^\perp$.

First, since $ X^*_{\tau^*}$ takes values in a compact subset, the weak convergence implies that:
\begin{align*}
\E[X^*_{\tau^*}] &= \int x \mu(dx,dz)= \underset{n\rightarrow \infty}{\lim} \int x \mu^n(dx,dz)=X_0
\end{align*}
We next prove that $ X^*$ is independent of $ \sigma(B_{0 \le s \le \tau^* })$. By construction of $ X^*$, we have that $ 
\E[X^*_{\tau^*} | \sigma(B_{0 \le s \le \tau^* }) ]= \E[X^*_{\tau^*} | Z_{\tau^*} ]$, and so we have to prove that $\E[X^*_{\tau^*} | Z_{\tau^*} ]=X_0$, i.e. for all bounded continuous function $ \phi$:
$$ 
\E[(X^*_{\tau^*}-X_0)\phi(Z_{\tau^*})] = \int_{{\rm cl}(O)} (x-X_0)\phi(z)\mu(dx,dz)=0. 
$$
By continuity of $ \phi $, and the fact that $ \mu$ is compactly supported, we have that:
\begin{align*}
\int_{{\rm cl}(O)} (x-X_0)\phi(z)\mu(dx,dz)&= \underset{n\rightarrow +\infty}{\lim}  \int_{{\rm cl}(O)} (x-X_0)\phi(z)\mu^n(dx,dz)\\
&= \underset{n\rightarrow +\infty}{\lim} \E[(X^n_{\tau^n_{n+1}}-X_0)\phi(Z_{\tau^n_{n+1}})].
\end{align*}
We next compute that:
 \b*
 \E\big[(X^n_{\tau^n_{n+1}}-X_0)\phi\big(Z_{\tau^n_{n+1}}\big)] 
 &=& 
 \E \Big[ \Big(\sum_{i=1}^{n+1} X^n_{\tau^n_i}-X^n_{\tau^n_{i-1}} \Big) 
              \phi\big(Z_{\tau^n_{n+1}}\big) \Big] 
 \\
 &=& 
 \sum_{i=1}^{n+1}  \E \big[ \E_{\tau^n_i} \big\{ \big( X^n_{\tau^n_i}-X^n_{\tau^n_{i-1}} \big)
                                                                         \phi\big(Z_{\tau^n_{n+1}}\big) 
                                                                \big\} 
                                   \big] 
 \\
 &=&  
 \sum_{i=1}^{n+1}  \E \big[ \big( X^n_{\tau^n_i}-X^n_{\tau^n_{i-1}} \big) 
                                          \E_{\tau^n_i } \big\{  \phi\big(Z_{\tau^n_{n+1}}\big) \big\} 
                                  \big].
 \e*
By continuity of $ Z$, we have that $ \E_{\tau^n_i } \big[  \phi\big(Z_{\tau^n_{n+1}}\big) \big]= \E_{\tau^n_i- } \big[  \phi\big(Z_{\tau^n_{n+1}}\big) \big] $, and therefore:
 \b*
 \E \big[ \big( X^n_{\tau^n_i}-X^n_{\tau^n_{i-1}} \big) 
             \E_{\tau^n_i } \big\{  \phi\big(Z_{\tau^n_{n+1}}\big) \big\} \big] 
 &=& 
 \E \big[ \big( X^n_{\tau^n_i}-X^n_{\tau^n_{i-1}} \big) 
              \E_{\tau^n_i- } \big\{  \phi\big(Z_{\tau^n_{n+1}}\big) \big\} \big]  
 \\
 &=&
 \E \big[ \E_{\tau^n_i- } \big\{  \phi\big(Z_{\tau^n_{n+1}}\big) \big\}  
             \E_{\tau^n_{i}-} \big\{ X^n_{\tau^n_i}-X^n_{\tau^n_{i-1}}  \big\}    
     \big] 
 \;=\;
 0,
 \e*
where we used the fact that $ \E_{\tau^n_i-}\big[ X^n_{\tau^n_i}\big]=X^n_{\tau^n_{i-1}}$. This concludes the proof.\\

\textit{Step 4:} We finally prove that we have $ (X^*+R(Z^z))_{\cdot \wedge \tau^*} \geq 0$ and $  \lbrace U(X^*_{\tau^* \wedge \theta}+R(Z^z_{\tau^* \wedge \theta}))^- \rbrace_{\theta \in \mathcal{T}}$ is uniformly integrable. We shall in fact prove that $ X^*_{\cdot \wedge \tau^*}+R(Z^z_{\cdot \wedge \tau^*}) >C$ $ dt \times d\P$-a.e. for some $ C>0$, which ensures that the two previous properties are trivially verified.\\

For that purpose, first notice that since ${{\rm cl}(O)}$ is a compact subset of $ int(\bar{D})$, we have the existence of $ C>0$ such that 
$$\P(X^*_{\tau^*}+R(Z_{\tau^*}) >C) =1,$$
that is equivalent to:
\begin{align}\label{eq: last interm}
 \E \Big[ \big( X^*_{\tau^*}+R(Z_{\tau^*})\big) {\bf{1}}_{A}\Big]>C \P[A], \  \text{for all} \ A \ \mathcal{F}_{\tau^*}-\text{measurable}.  
\end{align}
We now show that $ \P[x+ R(Z_{\tau^*}) >C]=1$. Indeed we have $ x+ R(Z_{\tau^*})= (X^*+Z)_{\tau^*-} $. Then assume that we have $ \epsilon >0$ such that $ \P \left[ A \right]>0$, where $ A := \lbrace  x+ R(Z_{\tau^*}) < C-\epsilon\rbrace$ . Using the fact that $ X^*$ is a martingale, we obtain:
$$ \E \Big[ \big( X^*_{\tau^*}+R(Z_{\tau^*}\big){\bf{1}}_{A} \Big] = \E \Big[ \big( x+ R(Z_{\tau^*}) \big){\bf{1}}_{A} \Big]< (C-\epsilon)\P[A],$$
which contradicts \ref{eq: last interm}.\\

Finally we recall from the choice of $ \tau^*$ in Step 2 that $ \tau^*$ is smaller than the exit time of the support of $ \mu_z$, which means in particular that we have $ Z_{\cdot\wedge \tau^*} \geq S(C-x) $ $ dt\times d\P$-a.s.
\ep

\section{Appendix: power utility function}
\label{sect:example}

Our goal is to compute explicitly the function $ \bar{U}^\infty$ in the context of the power utility function of Section \ref{sect:puc}. Proposition \ref{calcul.expl} then follows immediately from our explicit calculations.

The scale function $ S_\gamma$ of $ Y$ is given up to an affine transformation by
$$ S_\gamma(y)=\text{sgn}(1-\gamma)y^{1-\gamma} \ \ \text{if} \ \gamma \neq 1 \  \text{and} \ S_1(y)=\ln(y).$$
Then, the corresponding inverse function is:
 \b*
 R_\gamma(z)
 :=
 \left(\text{sgn}(1-\gamma)z \right)^{\frac{1}{1-\gamma}} 
 &\text{whenever}~~\text{sgn}(1-\gamma)z\in \R_+ ,&
 \mbox{for}~~\gamma \neq 1,
 \e*
and
 \b*
 R_1(z) := e^z 
 &\text{for all}& 
 z\in \R.
 \e*
The process $ Z$ is the martingale defined by:
 \b* 
 Z_t
 :=
 Z_0 e^{|1-\gamma|\sigma B_t-\frac{1}{2}(1-\gamma)^2\sigma^2 t},
 &\mbox{with}&
 Z_0=\text{sgn}(1-\gamma) Y_0^{1-\gamma}, 
 ~~\text{for}~~\gamma \neq1,
 \e*
and
 \b* 
 Z_t := Z_0+\sigma B_t, 
 &\mbox{with}&
 Z_0=\ln(Y_0),~~\text{for}~~\gamma=1.
 \e*
For notational convenience, we will drop the dependance of $R$ on $ \gamma$. 

\vspace{5mm} 

\no\textbf{Proof of Proposition \ref{calcul.expl}} We consider separately several cases.
\\
(i) $ \gamma <1$: Then, the domain of $ R$ is $ (0,+\infty)$.
          \\
          (i-1) $ p \neq 1$: We first recall the value of the derivatives with respect to $ z$:
                 $$ \partial_z \bar{U}(x,z)=\frac{1}{1-\gamma}z^{\frac{\gamma}{1-\gamma}} \left(x+z^{\frac{1}{1-\gamma}} \right)^{-p} $$
                  $$ \partial_{zz} \bar{U}(x,z)= \frac{1}{(1-\gamma)^2}z^{\frac{2\gamma-1}{1-\gamma}}\left(x+z^{\frac{1}{1-\gamma}} \right)^{-p-1} \left[\gamma \left(x+z^{\frac{1}{1-\gamma}} \right)-pz^{\frac{1}{1-\gamma}}  \right]  $$
                         \\
                         (i-1a) $ \gamma >p$: For any $ x $, $ \partial_{zz}\bar{U} (x,z)>0$ for $ z$ large enough. Since the domain of this function is $ (0,\infty)$, and $ \bar{U}(x,z)\rightarrow+\infty$ when $ z\rightarrow +\infty$, we have $ \bar{U}^1(x,\cdot)=+\infty$. So $ \bar{U}^\infty=\bar{U}^1=+\infty$.
                          \\
                          (i-1b) $ \gamma =p$:
                                 For $ x >0$, $ \partial_{zz}\bar{U} (x,z) >0$ and the same argument as above leads to $ \bar{U}^1(x,z)=+\infty$.
                                 For $ x \le 0$, $ \partial_{zz}\bar{U} (x,z) \le 0$ and then $ \bar{U}^1(x,z)=\bar{U}(x,z)$.\\
                                 We then have $ \bar{U}^1(x,z)= \bar{U}(x,z) {\bf{1}}_{x \le 0}+\infty  {\bf{1}}_{x > 0} $. For $ z \in (0,\infty)$, we now focus on the function $ \bar{U}^1(\cdot,z)$ on $ (-z^{\frac{1}{1-\gamma}},\infty)$. Since $ \bar{U}^1=+\infty$ for $ x$ large enough, we have $ \bar{U}^2(x,z)=+\infty$ for every $ (x,z)$ in the domain. So $ \bar{U}^\infty=\bar{U}^2=+\infty$
                            \\
                          (i-1c) $ \gamma <p$:
                                        \\
                                        $\bullet$ $\gamma \le 0$ leads to $ \partial_{zz}\bar{U} (x,z) \le 0$ so that $ \bar{U}$ is concave w.r.t. $ x$ and $ z$ and then $ \bar{U}^\infty=\bar{U}$.
                                        \\
                                        $\bullet$ $ \gamma >0$.
                                        For $ x \le 0$, we have $ \partial_{zz}\bar{U} (x,z) \le 0$ so that $ \bar{U}^1(x,\cdot)=\bar{U}(x,\cdot)$.
                                        For $ x>0$, there exists $ z(x)$ such that $ \partial_{zz}\bar{U} (x,z)>0$ for $ z<z(x)$ and $ \partial_{zz}\bar{U} (x,z)\le 0 $ for $z\geq z(x)$. Since $ \partial_{z}\bar{U} (x,z)\rightarrow 0 $ when $ z \rightarrow +\infty$, there exists $ \tilde{z}(x)$ such that $ \bar{U}^1(x,z)=U(x)+z \partial_z \bar{U}(x,\tilde{z}(x))$ for $ z \le \tilde{z}(x)$ and $ \bar{U}^1(x,z)=\bar{U}(x,z)$ for $ z >\tilde{z}(x)$. We see that $ z(x) $ is the unique solution of:
                                        $$ \bar{U}(x,z(x))-U(x)=z(x) \partial_z \bar{U}(x,z(x)).$$
                                        i.e. if we denote $ \xi(x):=x^{-1}z(x)^{\frac{1}{1-\gamma}}$, then $ \xi(x)$ is the unique solution of $ \Theta(\xi)=0$ where:
                                        $$ \Theta(\xi):=\frac{\left(1+\xi\right)^{1-p}-1}{1-p}-\frac{\xi}{1-\gamma} \left(1+\xi \right)^{-p}. $$
                                        We easily observe that $ \xi_0:=\xi(x)$ is independant of $ x$ and then:
                                        $$ \bar{U}^1(x,z)=\bar{U}(x,z) {\bf{1}}_{x\xi_0 \le z^{\frac{1}{1-\gamma}}}+\left(\frac{x^{1-p}-1}{1-p}+z x^{\gamma-p}\frac{\xi_0^\gamma}{1-\gamma} (1+\xi_0)^{-p} \right){\bf{1}}_{x\xi_0 > z^{\frac{1}{1-\gamma}}}.$$
                                        
                                        Notice that $ \partial_{xx} \bar{U}^1(x,z) \le 0$ on $ (-z^{\frac{1}{1-\gamma}},\frac{z^{\frac{1}{1-\gamma}}}{\xi_0})$. On the interval $ (\frac{z^{\frac{1}{1-\gamma}}}{\xi_0},+\infty)$, we compute that:
                                        $$ \partial_x\bar{U}^1(x,z)=x^{-p}+\frac{\gamma-p}{1-\gamma}x^{\gamma-p-1} z \xi_0^\gamma(1+\xi_0)^{-p} $$
                                        $$ \partial_{xx}\bar{U}^1(x,z)=-px^{-p-1}\left[1-\frac{(\gamma-p)(\gamma-p-1)}{p(1-\gamma)}z x^{\gamma-1} \xi_0^\gamma (1+\xi_0)^{-p}  \right].$$
                                        
                                        In order to investigate the sign of $ \partial_{xx} \bar{U}^1$, we introduce the function 
                                         \b*
                                         \Delta(\xi):=1-\frac{(p+1-\gamma)(p-\gamma)}{p(1-\gamma)} \xi_0^\gamma \xi^{1-\gamma}(1+\xi_0)^{-p}, & \xi \in [0,\xi_0],&
                                         \e* 
                                         and search for a solution $ \xi_1 $ to the equation $ \Delta(\xi)=0$.
                                         
The function $ \Delta$ is non-increasing with  $ \Delta(0)=1$. In order to investigate the sign of $ \Delta(\xi_0)$, we introduce the function:
 \b*
 \tilde{\Delta}(x)
 :=
 1 -\frac{(p+1-\gamma)(p-\gamma)}{p(1-\gamma)} x (1+x)^{-p} \;>\; 0
 &\mbox{for all}&
 x>0.
 \e*
This is clearly a non-increasing continuous and one-to-one function on $ (0,\infty)$, and we see that the sign of $ \Delta(\xi_0)$ reduces to the sign of  $ \tilde{\Delta}(x)$ under the condition $ \Theta(x)=0$. We are then reduced to the system of non-linear equations:
\be \label{deltasystem}
\tilde{\Delta}(x) =0
&\mbox{and}&
\Theta(x) =0,
\ee
which is immediately seen to be equivalent to:
\begin{align*}
& (1+\xi_0)^{-p} =\frac{1-\gamma}{1+p-\gamma} \\
& 1+ \frac{(1+\xi_0)^{-p}}{1-\gamma} \left[\left( \gamma-p\right)\xi_0-\left(1-\gamma \right) \right]=0
\end{align*}
We can see after calculus that the solution of \eqref{deltasystem} is $ x=\frac{p}{p-\gamma}$. Moreover, for a fixed  $ p$, we have:
$$ 
G(\gamma)=0 \Leftrightarrow \ \text{there is a unique solution to \reff{deltasystem},}
$$
where
$$ G(\gamma):=(p-\gamma)^p(p+1-\gamma)-(2p-\gamma)^p(1-\gamma).$$
Since $ G$ is a non-decreasing continuous and one-to-one function, it admits a unique solution $ \hat{\gamma}_p$. Moreover, we have that $ G$ is negative on $ \gamma \le \hat{\gamma}_p$ and positive on $ \gamma > \hat{\gamma}_p$. This result gives us that:
\\
$\star$ For $ \gamma > \hat{\gamma}_p$, $ G$ positive implies $ \tilde{\Delta}(x)$ negative. It means that $ \Delta(\xi_0)$ is negative, so $ \bar{U}^1$ is not concave in its first variable and admits an inflexion point to be determined.\\
$\star$ For $ \gamma \le \hat{\gamma}_p$, $ G$ negative implies $ \tilde{\Delta}(x)$ positive. This means that $ \Delta(\xi_0)$ is positive, so $ \bar{U}^1$ is concave in its first variable.

                                        We now focus on the case $ \gamma > \hat{\gamma}_p$. We are looking for a pair $ (x_1,x_2)$ such that $ x_1 \le \frac{z^{\frac{1}{1-\gamma}}}{\xi_0} < x_2$ and $ x_1$ maximal such that:
                                        \begin{align}\label{system of eq}
                                           \frac{\bar{U}^1(x_2,z)-\bar{U}^1(x_1,z)}{x_2-x_1} = \partial_x \bar{U}^1(x_2,z) \le \partial_x \bar{U}^1(x_1,z).
                                        \end{align}
                                            This is the characterization of the concave envelope of $ \bar{U}^1$ w.r.t. $ x$. We observe that this pair exists since $ \partial_x \bar{U}^1(x,z)\rightarrow 0$ when $ x \rightarrow +\infty$ and $ \partial_x \bar{U}^1(x,z) \rightarrow +\infty$ when $ x \rightarrow -z^{\frac{1}{1-\gamma}}$.\\
                                            An other remark is that for any $ \lambda >0$, we have $ \frac{\bar{U}^1(\lambda x_2,\lambda^{1-\gamma}z)-\bar{U}^1(\lambda x_1,\lambda^{1-\gamma}z)}{\lambda x_2-\lambda x_1}=\lambda^{-p} \frac{\bar{U}^1(x_2,z)-\bar{U}^1(x_1,z)}{x_2-x_1}$ and $ \partial_x \bar{U}^1(\lambda x_i ,\lambda^{1-\gamma} z)=\lambda^{-p} \partial_x \bar{U}^1(x_i,z)$ for $ i\in \lbrace 1,2 \rbrace$. We then see that there exists $ \xi_1$ and $ \xi_2$ such that for any $ (x,z)\in int(\bar{D})$, we have $ (x_1,x_2)=(\frac{z^{\frac{1}{1-\gamma}}}{\xi_1},\frac{z^{\frac{1}{1-\gamma}}}{\xi_2}) $.
                                            
                                            Finally we can compute the value of $ \bar{U}^2$:
                                            \begin{align*}
                                            \bar{U}^2(x,z)= &\bar{U}(x,z){\bf{1}}_{x\xi_1 \le z^{\frac{1}{1-\gamma}}}+\bar{U}^1(x,z){\bf{1}}_{x\xi_2 \geq z^{\frac{1}{1-\gamma}}} + \left( \bar{U}^1\left( \frac{z^{\frac{1}{1-\gamma}}}{\xi_2},z \right)  \right.\\
                                            & \left.+\left(x- \frac{z^{\frac{1}{1-\gamma}}}{\xi_2}\right)\partial_x \bar{U}^1\left( \frac{z^{\frac{1}{1-\gamma}}}{\xi_2},z \right) \right) {\bf{1}}_{ \frac{z^{\frac{1}{1-\gamma}}}{\xi_1}< x \le \frac{z^{\frac{1}{1-\gamma}}}{\xi_2}}. 
                                            \end{align*}
                                            By construction, $ \bar{U}^2$ is concave w.r.t. $ x$. For the concavity w.r.t. $ z$, we already know that $ \partial_{zz}\bar{U}^2 \le 0$ out of  $ \left[ \left(x \xi_2 \right)^{1-\gamma},\left(x \xi_2 \right)^{1-\gamma} \right]$. We also obtain by tedious calculations that $ \partial_{zz}\bar{U}^2 \le 0$ on $ \left( \left(x \xi_2 \right)^{1-\gamma},\left(x \xi_2 \right)^{1-\gamma} \right)$, and that $ \partial_{z-} \bar{U}^2 \left(x,\left(x \xi_2 \right)^{1-\gamma} \right) \geq \partial_{x+} \bar{U}^2 \left( x,\left(x \xi_2 \right)^{1-\gamma} \right) $, and $ \partial_{z-} \bar{U}^2 \left(x,\left(x \xi_1 \right)^{1-\gamma} \right) \geq \partial_{z+} \bar{U}^2 \left(x, \left(x \xi_1 \right)^{1-\gamma} \right) $, where $ \partial_{z-}$ (resp $ \partial_{z+}$ ) corresponds to the left derivative (resp the right derivative) with respect to $ z$.
                                        \\                      
                 (i-2) $ p=1$: The derivatives w.r.t. $ z$ are:
                 $$ \partial_z\bar{U}(x,z) = \frac{1}{1-\gamma} z^{\frac{\gamma}{1-\gamma}} \left(x+z^{\frac{1}{1-\gamma}} \right)^{-1},$$
                 $$ \partial_{zz}\bar{U}(x,z)=\frac{1}{(1-\gamma)^2}z^{\frac{2\gamma-1}{1-\gamma}}\left(x+z^{\frac{1}{1-\gamma}} \right)^{-2} \left[\gamma\left(x+z^{\frac{1}{1-\gamma}} \right)- z^{\frac{1}{1-\gamma}}  \right]. $$
                          \\
                          (i-2a) $ \gamma \le 0$: In that situation $ \partial_{zz}\bar{U} \le 0$ and then $ \bar{U}^\infty=\bar{U}$.
                           \\
                           (i-2b) $ \gamma>0$: If $ x \le 0$, then $ \partial_{zz}\bar{U}(x,z) \le 0$ and $ \bar{U}^1(x,z)=\bar{U}(x,z)$.
                           
                             If $ x>0$, there is an inflection point, similarly to the case $ \gamma <p$, $ p \neq 1$. We find $ z(x)$ such that $ \partial_{zz}\bar{U} (x,z)>0$ for $ z<z(x)$ and $ \partial_{zz}\bar{U} (x,z)\le 0 $ for $z\geq z(x)$. Since $ \partial_{z}\bar{U} (x,z)\rightarrow 0 $ when $ z \rightarrow +\infty$, there exists $ \tilde{z}(x)$ such that $ \bar{U}^1(x,z)=U(x)+z \partial_z \bar{U}(x,\tilde{z}(x))$ for $ z \le \tilde{z}(x)$ and $ \bar{U}^1(x,z)=\bar{U}(x,z)$ for $ z >\tilde{z}(x)$. We see that $ z(x) $ is the unique solution of:
                                        $$ \bar{U}(x,z(x))-U(x)=z(x) \partial_z \bar{U}(x,z(x)).$$
                                        i.e. if we denote $ \xi(x):=x^{-1}z(x)^{\frac{1}{1-\gamma}}$, then $ \xi(x)$ is the unique solution of:
                                        $$ \ln\left(1+\xi \right)=\frac{\xi}{1-\gamma} \left(1+\xi \right)^{-1}. $$
                                        We easily observe that $ \xi_0:=\xi(x)$ is independant of $ x$ and then:
                                        $$ \bar{U}^1(x,z)=\bar{U}(x,z) {\bf{1}}_{x\xi_0 \le z^{\frac{1}{1-\gamma}}}+\left(\ln(x)+z x^{\gamma-1}\frac{\xi_0^\gamma}{1-\gamma} (1+\xi_0)^{-1} \right){\bf{1}}_{x\xi_0 > z^{\frac{1}{1-\gamma}}}.$$
                                        The derivation of $ \bar{U}^2$ is similar to the previous case. Indeed, for $ x \le \frac{z^{\frac{1}{1-\gamma}}}{\xi_0} $, $ \partial_{xx} \bar{U}^1(x,z) \le 0 $ by definition of $ U$.\\
                                        For $ x \geq \frac{z^{\frac{1}{1-\gamma}}}{\xi_0} $, we have:
                                        $$ \partial_x \bar{U}^1(x,z)= \left[ x^{-1}-z x^{\gamma-2} \xi_0^\gamma(1+\xi_0)^{-1} \right],$$
                                        $$ \partial_{xx} \bar{U}^1 (x,z) =-x^{-2} \left[1+(2-\gamma) z x^{\gamma-1} \xi_0^\gamma (1+\xi_0)^{-1} \right].$$
                                        The exact same scheme as the one leading to the system of equations \eqref{deltasystem} leads to the existence of $ \hat{\gamma}_1\in (0,1) $ such that for $ \gamma \le \hat{\gamma}_1$, we have $ \partial_{xx}\bar{U}^1 \le 0$, and for $ \gamma > \hat{\gamma}_1$, there exists an inflexion point.
                                        
                                        It remains to solve the case $ \gamma > \hat{\gamma}_1$. We are seeking for a pair $ (x_1,x_2)$ such that $ x_1 \le \frac{z^{\frac{1}{1-\gamma}}}{\xi_0} < x_2$ with $ x_1$ maximal such that \eqref{system of eq} is true. By the same arguments, there exists $ \xi_1$ and $ \xi_2$ such that for any $ z >0$, we have $ (x_1,x_2)=\left( \frac{z^{\frac{1}{1-\gamma}}}{\xi_1},\frac{z^{\frac{1}{1-\gamma}}}{\xi_2}\right) $ and:
                                        \begin{align*}
                                            \bar{U}^2(x,z)= &\bar{U}(x,z){\bf{1}}_{x\xi_1 \le z^{\frac{1}{1-\gamma}}}+\bar{U}^1(x,z){\bf{1}}_{x\xi_2 \geq z^{\frac{1}{1-\gamma}}}\\
                                            &+\left( \bar{U}^1\left( \frac{z^{\frac{1}{1-\gamma}}}{\xi_2},z \right)+\left(x- \frac{z^{\frac{1}{1-\gamma}}}{\xi_2}\right)\partial_x \bar{U}^1\left( \frac{z^{\frac{1}{1-\gamma}}}{\xi_2},z \right) \right) {\bf{1}}_{ \frac{z^{\frac{1}{1-\gamma}}}{\xi_1}< x \le \frac{z^{\frac{1}{1-\gamma}}}{\xi_2}}. 
                                            \end{align*}
                                            The concavity in $ z$ is easily obtained by direct calculations.
 \\ 
 (ii) $ \gamma =1$: The admissible domain of $ R$ is $ (-\infty,\infty)$.
\\
(ii-1) $ p\neq 1$: We have:
                       $$ \partial_{x}\bar{U}(x,z)=e^z\left(x+e^z \right)^{-p},  $$
                       $$ \partial_{xx}\bar{U}(x,z)=e^z \left( x+ e^z \right)^{-p-1} \left[\left(x+e^z\right)-pe^z \right].$$
\\
(ii-1a) $ p<1$: If $ x \geq 0$, then $ \partial_{zz}\bar{U}(x,z) > 0$ and then since $ z$ is unbounded ($ \forall z \in \R$, $ x+e^z >0 $ if $ x \geq 0$), and $ \bar{U}(x,\cdot)$ is strictly convex and $ \bar{U}(x,z) \rightarrow +\infty$ when $ z \rightarrow +\infty$, we have $ \bar{U}^1(x,z)=+\infty$.

                           For $ x<0$, we have $ \partial_{zz}\bar{U}(x,z) \le 0$ for $ z \le \ln \left( \frac{1-p}{x} \right)$ and $ \partial_{zz}\bar{U}(x,z)>0$ for $ z>\left( \frac{1-p}{x} \right) $, and the same argument leads to $ \bar{U}^1(x,z)=+\infty$.\\
(ii-1b) $ p>1$: If $ x \le 0$, then $ \partial_{zz}\bar{U}(x,z) \le 0$ and $ \bar{U}^1(x,z)=\bar{U}(x,z)$.
                           For $ x>0$, we have $ \partial_{zz}\bar{U}(x,z) >0$ for $ z < \ln \left( \frac{x}{p-1} \right)$ and $ \partial_{zz}\bar{U}(x,z) \le 0 $ for $ x \geq \ln \left( \frac{x}{p-1} \right)$ . Since $ \bar{U}(x,z) \rightarrow U(x)>-\infty $ when $ z \rightarrow -\infty$, and $ \bar{U}(x,z) \rightarrow -\frac{1}{1-p}$ when $ z\rightarrow +\infty$, we have that the concave envelope is always equal to the limit when $ z\rightarrow +\infty$, i.e. $ \bar{U}^1(x,z)= \frac{1}{p-1}$. So:
                           $$ \bar{U}^1(x,z)= \bar{U}(x,z) {\bf{1}}_{x \le 0}+\frac{1}{p-1} {\bf{1}}_{x>0}.$$
                           In particular we see that $ \bar{U}^1$ is not continuous.
                           
                           The calculation of $ \bar{U}^2$ is easier than in the previous cases. For a fixed $ z\in \R$. We study $ \bar{U}^1(\cdot,z)$ on $ \left(-e^z,\infty \right)$. $  \bar{U}^1(\cdot,z)$ is non decreasing, constant on $ [0,\infty)$ and concave on $ \left(-e^z,0 \right)$, with $ \bar{U}^1\left(-e^z,z\right)=-\infty$. So there exists $ x_0\in \left( -e^z,0 \right)$ such that $ \partial_x \bar{U}^1(x_0,z)=\frac{\bar{U}^1(0,z)-\bar{U}^1(x_0,z)}{-x_0} $, and $ \bar{U}^2(\cdot,z)$ is linear on $ (-x_0,0)$ and $ \bar{U}^2(x,z)=\bar{U}^1(x,z)$ elsewhere. $ x_0$ is easily given by $ x_0=-\frac{e^z}{p}$ and then:
                           \begin{align*}
                           \bar{U}^2(x,z)= &\bar{U}(x,z) {\bf{1}}_{x \le -\frac{e^z}{p}}-\frac{1}{1-p} {\bf{1}}_{x\geq 0}\\
                           &+\left( \bar{U}\left(-\frac{e^z}{p},z \right)+\left( x+ \frac{e^z}{p} \right)e^{-pz} \left( 1- \frac{1}{p} \right)^{-p} \right) {\bf{1}}_{x\in \left(-\frac{e^z}{p},0 \right)}.
                           \end{align*}
                           The partial concavity w.r.t. $ z$ is then trivial and we have $ \bar{U}^\infty=\bar{U}^2$.
\\
(ii-2) $ p=1$: we have:
                    $$ \partial_z \bar{U}(x,z)=\left(1+ x e^{-z} \right)^{-1}, $$
                    $$ \partial_{zz} \bar{U}(x,z)=xe^{-z}\left( 1+xe^{-z} \right)^{-2}.$$
                    For $ x>0$, we have $ \partial_{zz} \bar{U}(x,z)>0 $ and then as above, since $ \bar{U}(x,z)\rightarrow \infty$ when $ z\rightarrow \infty$, we have $ \bar{U}^1(x,z)=\infty$.\\
                    For $ x \geq 0$, we have $ \partial_{zz} \bar{U}(x,z) \le 0$ and then $ \bar{U}^1(x,z)=\bar{u}(x,z)$. Summing up:
                    $$ \bar{U}^1(x,z)=\bar{U}(x,z) {\bf{1}}_{x \le 0}+\infty{\bf{1}}_{x>0}.$$
                    As a consequence, we see that:
                    $$ \bar{U}^2=+\infty.$$
\\
(iii) $ \gamma >1$: The admissible domain of $ R$ is $ (-\infty,0)$.
            For any $ p$, the partial derivatives w.r.t. $ z$ are given by:
                 $$ \partial_z \bar{U}(x,z) = \frac{1}{\gamma-1} \left(-z \right)^{\frac{\gamma}{1-\gamma}} \left( x+\left(-z \right)^{\frac{1}{1-\gamma}} \right)^{-p},$$
                 $$ \partial_{zz} \bar{U}(x,z)=\frac{1}{(\gamma-1)^2} \left(-z \right)^{\frac{2\gamma-1}{1-\gamma}}\left(x+\left(-z \right)^{\frac{1}{1-\gamma}}\right)^{-p-1} \left[ \gamma \left( x+\left(-z \right)^{\frac{1}{1-\gamma}} \right)-p \left(-z \right)^{\frac{1}{1-\gamma}} \right]. $$
 \\
(iii-1) $ p \le 1$: For any x, $ \partial_{zz}\bar{U}(x,z) >0 $ for $ z $ large enough and $ \bar{U}(x,z) \rightarrow +\infty$ when $ z \rightarrow 0$ so that $ \bar{U}^1(x,z)=+\infty$. 
\\
(iii-2) $ 1<p <\gamma$: For $ x \geq 0$, we have $ \partial_{z}\bar{U}(x,z)\rightarrow 0 $ when $ z \rightarrow -\infty$ and $ \bar{U}(x,z) \rightarrow \frac{1}{p-1}$ when $ z \rightarrow 0$, so $ \bar{U}^1(x,z)=\frac{1}{p-1}$.\\
                    For $ x<0$, for $ z \le - \left(\frac{\gamma}{p-\gamma} x \right)^{1-\gamma}$, $ \partial_{zz}\bar{U}(x,z) \le 0 $ and for $ z >  - \left(\frac{\gamma}{p-\gamma} x \right)^{1-\gamma}$, $ \partial_{zz}\bar{U}(x,z) >0$. Since $ \bar{U}(x,z) \rightarrow \frac{1}{p-1} $ when $ z \rightarrow 0$, there exists $ z_0$ such that $-z_0 \partial_z \bar{U}(x,z_0)=\frac{1}{p-1}-\bar{U}(x,z_0) $. Similarly to the case $ \gamma <1$, $ z_0 $ verifies $ (-z_0)^{\frac{1}{1-\gamma}}=-x \xi_0$ with $ \xi_0 =\frac{\gamma-1}{\gamma-p} $.
                    
                    We then have:
                    \begin{align*}
                    \bar{U}^1(x,z)=&\bar{U}(x,z) {\bf{1}}_{ \lbrace -x \xi_0 > \left(-z\right)^{\frac{1}{1-\gamma}} \rbrace }+\frac{1}{p-1} {\bf{1}}_{\lbrace x\geq 0 \rbrace} \\
                    &+z \left(-x \right)^{\gamma-p} \frac{(p-1)^{-p}}{(\gamma-p)^{\gamma-p}} (\gamma-1)^{\gamma-1}  {\bf{1}}_{ \lbrace 0< -x \xi_0 \le \left(-z\right)^{\frac{1}{1-\gamma}} \rbrace }
                    \end{align*}
                   The concavity of $ \bar{U}^1$ w.r.t. $ x$ is then straightforward.
\\
(iii-3) $ p \geq \gamma $:  For $ x \le 0$, $ \partial_{zz} \bar{U}(x,z) \le 0$ and $ \bar{U}^1(x,z)=\bar{U}(x,z)$.\\
                 For $ x >0$, there is an inflexion point. Now since $ \partial_z \bar{U}(x,z)\rightarrow 0$ when $ z \rightarrow -\infty  $, we have $ \bar{U}^1(x,z)= \frac{1}{p-1}$. So:
                 $$ \bar{U}^1(x,z)=\bar{U}(x,z){\bf{1}}_{x \le 0}+ \frac{1}{p-1} {\bf{1}}_{x > 0}.$$
                 We now search $ \bar{U}^2$. For any $ z\in (-\infty,0)$, $ \bar{U}^1(\cdot,z)$ is concave on $ (-(-z)^{\frac{1}{1-\gamma}},0)$ and constant on $ [0,\infty)$, and discontinuous at $ x=0$. We are looking for $x_0 \in  (-(-z)^{\frac{1}{1-\gamma}},0) $ such that:
                 $$ \bar{U}^1(0,z)-\bar{U}(x_0,z)=-x_0 \partial_x \bar{U}(x_0,z).$$
                 The solution is given by $ x_0=\frac{1-p}{p} (-z)^{\frac{1}{1-\gamma}}$ and we have:
                 \begin{align*}
                 \bar{U}^2(x,z)=&\bar{U}(x,z) {\bf{1}}_{x <\frac{1-p}{p}(-z)^{\frac{1}{1-\gamma}}}+\frac{1}{p} {\bf{1}}_{x >0} \\
                 &+\left( (-z)^{\frac{1-p}{1-\gamma}}+\left( x+\frac{p-1}{p}(-z)^{\frac{1}{1-\gamma}}\right) p^p (-z)^{\frac{-p}{1-\gamma}} \right){\bf{1}}_{\frac{1-p}{p}(-z)^{\frac{1}{1-\gamma}} \le x <0}.
                 \end{align*}
                 The concavity of $ \bar{U}^2$ w.r.t. $ z$ is easily verified.
\ep

\end{document}